\documentclass[prb,aps,eqsecnum,preprint]{revtex4}
\usepackage{graphicx}

\def\be{\begin{equation}}
\def\ee{\end{equation}}
\begin{document}
%
%
\title{Depletion interactions of non-spherical colloidal particles \\ in polymer solutions}
\author{E. Eisenriegler and A. Bringer}
\affiliation{Institut f\"ur Festk\"orperforschung,
Forschungszentrum J\"ulich,
\\ D-52425 J\"ulich, Germany}

\date{\today}

\begin{abstract}
We consider anisotropic colloidal particles immersed in a solution
of long, flexible, and nonadsorbing polymers. For the dumbbell
shapes of recently synthesized particles consisting of two
intersecting spheres and for lens-shaped particles with spherical
surfaces we calculate the isotropic and anisotropic interaction
parameters that determine the immersion free energy and the
orientation-dependent depletion interaction between particles that
are induced by the polymers. Exact results are obtained for
random-walk like (ideal) polymer chains.
\end{abstract}

\maketitle
%
%
%

\section{INTRODUCTION}
\label{intro}

In colloidal suspensions containing polymer chains, there are
tunable effective interactions between the colloid particles. Free
nonadsorbing polymer chains avoid the space between two particles,
leading to an unbalanced pressure, which pushes them towards each
other. Such depletion forces for an isolated pair of immersed
particles or for a single immersed particle near a wall have been
measured in recent experiments \cite{rudh98}.

Here we consider {\it anisotropic} colloid particles. For the
dumbbell shapes of recently synthesized \cite{manoha} particles
consisting of two intersecting spheres, and for lens-shaped
particles with spherical surfaces, as in Fig. 1, we calculate the
immersion free energy and the {\it orientation-dependent}
depletion interaction. The predictions are compared with results
for prolate and oblate ellipsoids \cite{snoeblaa,ebm03,PCI}, which
have also a symmetry axis of revolution and a symmetry center of
reflection.

The case of {\it large} particle to polymer size ratio can be
investigated by means of small curvature expansions of the
Helfrich or Derjaguin type, but here we consider mesoscopic
particles which are {\it small} compared to characteristic polymer
lengths such as the gyration radius ${\cal R}_g$, and we
concentrate on the case of {\it ideal}, random-walk like,
polymers. The well known correspondence \cite{gen79} between the
statistics of long flexible polymers and critical field theories
allows us to use the {\it small particle operator-expansion}
\cite{be95,er95,eise04} for predicting the polymer-induced
interactions. The operator weights in the expansion for dumbbells
and lenses are calculated by a conformal mapping to a
wedge-geometry.

We introduce the polymer-magnet analogy and small particle
expansion in Sec. II, discuss density-profiles in a wedge and
outside a dumbbell or lens in Sec III, and evaluate, for ideal
polymers, the corresponding small particle amplitudes in the
Gaussian model in Sec IV. These results are compared with
corresponding results for ellipsoids and, in Sec. V, with a more
general class of weakly anisotropic particles. In Sec. VI the
amplitudes are used to determine the orientation-dependent
interactions, and in Sec VII we summarize the new results. Some
technical details are relegated to Appendices A-D.

\section{Polymer-magnet analogy and small particle
expansion}

In the polymer-magnet analogy the partition function of a polymer
chain with ends at ${\bf r}_1$ and ${\bf r}_2$ corresponds to the
order-parameter correlation function $\langle \varphi_{12}
\rangle$ of a Ginzburg-Landau model or field theory
\cite{PCI,gen79}. Here $\varphi_{12}$ is the product $\Phi({\bf
r}_1) \, \Phi({\bf r}_2)$ of two order parameter fields $\Phi$.
{\it Ideal} polymers correspond to a {\it Gaussian}
Ginzburg-Landau model with Hamiltonian
\be \label{Gaussham}
H \, = \, \int \,  d {\bf r} \, \biggl[ \frac{1}{2} ( \nabla \Phi
)^2 \, + \,  \frac{t}{2} \Phi^2 \biggr] \, ,
\ee
where the integration extends over the volume outside the
particles, and where the order parameter satisfies the Dirichlet
boundary condition
\be \label{diri}
{\Phi} = 0 \qquad
\ee
at the particle surfaces, since we consider nonadsorbing polymers.
We always consider length scales much larger than the persistence
and extrapolation lengths.

The free energy $F$ it costs to immerse particles in a dilute
polymer solution in an unbounded space or in the half space
bounded by a wall is determined by the polymer partition functions
with and without the particles and is given by \cite{ebm03,PCI}
\begin{eqnarray} \label{inter}
F/p_0 \, = \, - {\cal L} \, \int d{\bf r}_1  d{\bf r}_2 [\langle
\varphi_{12} \rangle_{{\cal H} + \delta {\cal H}} - \langle
\varphi_{12} \rangle_{{\cal H} }] \, .
\end{eqnarray}
Here $p_0 = n k_B T$ is the ideal gas pressure in the dilute
solution with chain density $n$, and ${\cal H} + \delta {\cal H}$
and ${\cal H}$ denote Ginzburg-Landau Hamiltonians of the form
(\ref{Gaussham}) in presence and absence of the particles,
respectively. The dependence of the double integral on the
temperature deviation $t$ from the critical point is converted
into the dependence on ${\cal R}_g^2$ of $-F/p_0$ by means of the
inverse Laplace transform ${\cal L}\, (..)=\int (d t / 2\pi i){\rm
exp}(3t{\cal R}_g^2 /d) \, (..)$ where $d$ denotes the spatial
dimension.

Consider particles with a size much smaller than ${\cal R}_g$ and
a shape that is symmetric about both a center of reflection and an
axis of revolution. Examples are rods, disks, ellipsoids,
dumbbells, and lenses. In the spirit of the operator-product
expansion, a small mesoscopic perturbation in a critical field
theory can be represented by a sum of point operators. Thus, for a
single small particle \cite{be95,er95,eise04,contact} with center
at ${\bf r}_{\rm P}$,

\begin{eqnarray} \label{small}
e^{- \delta {\cal H}} \, \propto \, 1 \, + \, \sigma_I \, + \,
\sigma_A \, ,
\end{eqnarray}
where
\begin{eqnarray} \label{small'}
\sigma_I \, = \, a_1 \, \epsilon({\bf r}_{\rm P}) \, + \, ... \, ,
\end{eqnarray}
\begin{eqnarray} \label{small''}
{\sigma}_A \, = \, b_1 \, \partial_{\parallel}^2 \epsilon({\bf
r}_{\rm P}) \, + \, b_2 \, T_{\parallel \parallel}({\bf r}_{\rm
P}) \, + \, ...
\end{eqnarray}
are linear combinations of {\it isotropic(I)} and {\it
anisotropic(A)} operators from the operator algebra of the
Ginzburg-Landau model, which reflect the symmetries of the
particle shape and boundary condition. Here $\epsilon \propto
-\Phi^2$ is the energy density, $\partial_{\parallel}$ is a
derivative along the particle axis, and $T_{{\parallel}
{\parallel}}$ is the diagonal component of the stress tensor of
the field theory along the axis.

Only the operators of lowest scaling dimensions $x=d-1/\nu, \,
d+2-1/\nu$, $d$ are shown in Eqs. (\ref{small})-(\ref{small''}),
and their coefficients $a_1, \, b_1, \, b_2$ by scale invariance
must be proportional to the particle size raised to the power $x$.
For ideal chains (Gaussian model) the Flory (correlation length)
exponent $\nu$ equals $1/2$, so that $\partial_{\parallel}^2
\epsilon$ and $T_{\parallel
\parallel}$ have the same scaling dimension $d$, and the $b_1$ and
$b_2$ terms both contribute to the leading anisotropic behavior.
For chains with excluded volume interaction (corresponding
\cite{gen79} to the $N$-vector model in the limit $N \to 0$),
$\nu$ is larger than $1/2$, and we expect that the $b_2$ term
dominates the anisotropic behavior of a small particle.

The coefficients $a_1, \, b_1, \, b_2$  depend on the size and
shape of the particle but are {\it independent} of other distant
particles, of the distant boundary wall of the half space, and
\cite{transl} of $t$ . Thus we evaluate the coefficients for a
{\it single} particle in an {\it unbounded} space at $t=0$, and
then use them to make predictions for the interaction between
particles or a particle and a wall. The coefficients can be
evaluated from the density profiles $\langle \epsilon ({\bf r})
\rangle$ and $\langle T_{kl} ({\bf r}) \rangle$ of the energy
density and the stress tensor that are induced by a single
particle. While ellipsoids have been considered in Refs.
\onlinecite{ebm03,PCI} we concentrate here on dumbbells and
lens-shaped particles.

\section{DENSITIES IN A WEDGE AND OUTSIDE A DUMBBELL OR LENS}

A system at the critical point containing a colloidal dumbbell
composed of two overlapping spheres, or a lens with two spherical
surfaces, can be conformally mapped onto a critical system filling
a wedge with opening angle $\alpha$, which is smaller or larger
than $\pi$ in case of the dumbbell or lens, as in Fig. 1. As
explained in more detail in Ref. \onlinecite{eise04}, an inversion
about the point denoted by the heavy dot on the left hand side of
Fig. 1 maps the interior of the wedge onto the exterior of a
particle with a dumbbell or lens shape. The two boundary half
planes of the wedge and the edge where they meet are mapped onto
the two spherical surfaces of the particle and the circle $C$ of
diameter ${\cal D}$ where they intersect. ${\cal D}$ is related to
the diameter $L$ of the two spheres by ${\cal D}=L {\rm
sin}(\alpha/2)$. For $\alpha=0, \pi$, and $2 \pi$ the particle on
the right hand side of Fig. 1 becomes a dumbbell of two touching
spheres, a spherical particle, and a circular disk, respectively.

For the wedge the boundary-induced density profile of a scalar
operator ${\cal O}$, such as the energy density $\epsilon$, has
the form \cite{cardy,be85,hksd99}
\be \label{scalwedge}
\langle {\cal O} (\hat{\bf r}_{{\rm e}} , \rho , \Omega)
\rangle_{\rm wedge}^{(i,j)} = B_{\cal O}^{1/2} \rho^{-x_{\cal O}}
{\bar {\cal P}}_{i, j} (\alpha , \Omega) \, .
\ee
Here the position vector $\hat{\bf r}$ is expressed in cylindrical
coordinates $(\hat{\bf r}_{{\rm e}} , \rho , \Omega)$, where the
edge of the wedge is the axis. The component $\hat{\bf r}_{\rm e}$
is parallel to the edge (and in general has dimension $d-2$), and
the two component vector perpendicular to the edge is determined
by its angle $\Omega$ with the symmetry half plane of the wedge,
i.e. $-\alpha /2 \le \Omega \le \alpha /2$, and by the distance
$\rho$ from the edge. The indices $(i,j)$ characterize the surface
universality classes \cite{edgenote} of the two boundary half
planes $\Omega=(-\alpha/2,\alpha/2)$ of the wedge. While ${\bar
{\cal P}}$ is a universal scaling function, $B_{\cal O}$ is the
non-universal amplitude in the pair correlation function
\be \label{engbulk}
\langle {\cal O}({\bf r}) {\cal O}(0) \rangle_{\rm bulk} =
B_{{\cal O}} \, r^{-2 x_{\cal O}}
\ee
of ${\cal O}$ in the bulk.

The corresponding density $\langle {\cal O}({\bf r}) \rangle_{\rm
particle}$ outside a dumbbell or lens follows \cite{cardy} from
(\ref{scalwedge}) and the conformal mapping and has the form given
in Eqs. (5.14) and (5.10) of Ref. \onlinecite{eise04}. Turning to
the energy density ${\cal O}=\epsilon$ with scaling function
$\bar{\cal P}=\bar{\cal E}$ and equal boundaries $i=j$, the
behavior
\be \label{smallomega}
{\bar {\cal E}}_{i,i} (\alpha , \Omega) = e_{0}^{(i)} (\alpha) [1
+ e_{2}^{(i)} (\alpha) \;  \Omega^2 / 2 + ...] \;
\ee
near the symmetry half plane $\Omega=0$ of the wedge determines
the profile $\langle \epsilon({\bf r}) \rangle_{\rm particle}$ far
from the dumbbell or lens and yields \cite{eise04}
\be \label{isopart}
{a}_{1}^{(i)} = {\cal D}^{x_{\epsilon}} \, e_{0}^{(i)} (\alpha) /
B_{\epsilon}^{1/2}
\ee
and
\begin{eqnarray} \label{aniso++part}
{b}_{1}^{(i)} = {a}_{1}^{(i)} \, \frac{{\cal D}^{2}}{8
x_{\epsilon}(x_{\epsilon} + 1)} \, \bigl( e_{2}^{(i)} (\alpha) -
x_{\epsilon} \bigr) \;
\end{eqnarray}
for two leading coefficients in the small particle expansion.

The boundary-induced density profile $\langle
T_{\kappa\lambda}(\hat{\bf r}) \rangle_{\rm wedge}$ of the stress
tensor in the wedge is given by
\be \label{stresswedge}
\langle T_{\kappa\lambda} (\hat{\bf r}_{{\rm e}} , \rho , \Omega)
\rangle_{\rm wedge}^{(i,j)} =  \rho^{-d} \, \tau_{i,j}(\alpha) \,
[\delta_{\kappa\lambda} - d \, u_{\kappa}^{(n)}(\Omega) \,
u_{\lambda}^{(n)}(\Omega) ] \, ,
\ee
where ${\bf u}^{(n)}(\Omega)$ is the unit vector normal to the
half plane $\Omega={\rm const}$ which contains $\hat{\bf r}$, as
in the left hand side of Fig. 2. Unlike (\ref{scalwedge}) there is
no nonuniversal amplitude in (\ref{stresswedge}), and the
$\rho$-exponent and the $\Omega$-dependence are trivial. Only the
variation of the universal amplitude $\tau$ with the opening angle
$\alpha$ of the wedge depends on the bulk universality class and
the surface classes $i,j$, and remains to be determined. The
stress tensor density (\ref{stresswedge}) has a vanishing trace
and obeys the continuity equation, as discussed in Appendix A.

We note two special cases: (i) For $\alpha \to \pi$ and $i=j$ the
wedge becomes the half space with a uniform boundary, and $\tau
\to 0$ since the stress tensor density vanishes \cite{cardy}.
According to the left hand side of Fig. 2, only for $\tau=0$ is
(\ref{stresswedge}) consistent with the symmetries of the half
space. (ii) For $\alpha \to 0$, $\tau$ diverges as
\be \label{film}
\tau_{i,j} \, \to \, \alpha^{-d} \, (-\Delta_{i,j}) \, ,
\ee
where $\Delta_{i,j}$ is the universal amplitude which determines
the stress tensor density in the parallel plate geometry
\cite{cardy}. If the width of the film is $\omega$ and both tensor
components are parallel to the plates, $\langle T_{\rm parallel,
parallel} \rangle_{\rm film}^{(i,j)}=\omega^{-d} (-\Delta_{i,j})$.

Using the inversion transformation for the conformal stress tensor
\cite{cardyfilm,eise04} one finds from Eq. (\ref{stresswedge}) the
stress tensor density
\be \label{stresspart}
\langle T_{kl} ({\bf r}) \rangle_{\rm particle}^{(i,j)} = ({\cal
D}/\Lambda^{2})^d \, \tau_{i,j}(\alpha) \, [ \delta_{kl} -d \,
u_{k}^{(N)}({\bf r}) u_{l}^{(N)}({\bf r}) ]
\ee
outside the dumbbell or lens. Here
\be \label{Cdistance}
\Lambda^2 = \sqrt{[ r^2 - ({\cal D}/2)^2 ]^2 + {\cal D}^2
r_{\parallel}^2} \,
\ee
with $r$ and $r_{\parallel}$ the distance of point ${\bf r}$ from
the particle center and its component parallel to the particle
rotation axis. The inverse length ${\cal D}/\Lambda^2$ in
(\ref{stresspart}) equals $b(\hat{\bf r})/\rho(\hat{\bf r})$,
where $b$ is the dilatation factor $|{\rm det}(\partial \hat{\bf
r} /
\partial {\bf r})|^{1/d}$ of the conformal mapping, and diverges as
$(r_{\parallel},r) \to (0,{\cal D}/2)$, as the circle $C$ of
intersection is approached. The unit vector ${\bf u}^{(N)}$ in the
particle geometry is the counterpart of ${\bf u}^{(n)}$ in the
wedge geometry. It points along the surface normal at ${\bf r}$ of
the spherical surface portion $S_{C,{\bf r}}$ which contains ${\bf
r}$ and is bounded by the circle $C$. $S_{C,{\bf r}}$ is the image
of the half plane $\Omega=$ const which contains $\hat{\bf r}$.
With the particle axis $\parallel$ as one of the Cartesian
directions,
\be \label{uN}
u_{k}^{(N)}({\bf r}) = \Lambda^{-2} \, \{\delta_{k {\parallel}}
[r^2 - ({\cal D}/2)^2 ] - 2 r_k r_{\parallel} \} \, .
\ee
The vector field ${\bf u}^{(N)}$ for given ${\cal D}$ is shown on
the right hand side of Fig. 2. Both ${\bf u}^{(N)}$ and ${\bf
u}^{(n)}$ are {\it independent} of $\alpha$ and the bulk and
surface universality classes $i,j$.

In leading order ${\cal D} \ll r$,
\be \label{stressdist}
\langle T_{kl} ({\bf r}) \rangle_{\rm particle}^{(i,j)} \to ({\cal
D}/r^{2})^d \, \tau_{i,j}(\alpha) \, [ \delta_{kl} -d \, {\cal
I}_{k,{\parallel}} \, {\cal I}_{l,{\parallel}} ] \, ,
\ee
with
\be \label{I}
{\cal I}_{k,m} \equiv {\cal I}_{k,m}({\bf r}) \, = \, \delta_{k,m}
- 2 r_k r_m / r^2 \, .
\ee
On comparing with the stress tensor correlation function
\be \label{stress2}
\langle T_{k l}({\bf r}) T_{m n}(0) \rangle_{\rm bulk} = B_T \,
r^{-2d} \{ (1/2) [{\cal I}_{k,m}{\cal I}_{l,n}+{\cal I}_{k,n}{\cal
I}_{l,m}] - (1/d) \delta_{k l} \delta_{m n} \}
\ee
in unbounded bulk \cite{cardyfilm,mao95,eise04}, Eq.
(\ref{stressdist}) implies that
\be \label{larger}
\langle T_{k l}({\bf r}) \rangle_{\rm particle}^{(i,j)} \, \to \,
b_2^{(i,j)} \langle T_{k l}({\bf r}) \, T_{\parallel \,
\parallel}(0) \rangle_{\rm bulk} \quad , \quad {\cal D} \ll r \, ,
\ee
with the stress tensor contribution in the small particle
expansion given by
\be \label{b2}
b_2^{(i,j)} T_{\parallel \, \parallel}=-{\cal D}^d \tau_{i,
j}(\alpha) \frac{d}{B_T} T_{\parallel \, \parallel} \, .
\ee
For the special case $\alpha \to 0$, where ${\cal D} \to \alpha
L/2$ and Eq. (\ref{film}) applies, Eq. (\ref{b2}) reduces to the
expression $b_2^{(i,j)} \to (L/2)^d \Delta_{i, j} d/B_T $ for a
dumbbell of two touching spheres with diameter $L$, given in Eq.
(2.15) of Ref. \onlinecite{eise04}.

The form of $\langle T \rangle_{\rm wedge}^{(i,j)}$ in
(\ref{stresswedge}) has been calculated for special cases. See
Refs. \onlinecite{cardy,buxu91} for $d=2$, Ref.
\onlinecite{hksd99} for symmetry breaking surfaces $i=j$ as $d \to
4$, and Appendix B for Dirichlet boundaries and $\alpha=\pi/2$.
Here we indicate how (\ref{stresswedge}) can be derived in the
general case. For ${\bf r}$ on the particle rotation axis,
$\langle T \rangle_{\rm particle}^{(i,j)}$ must have an
eigenvector parallel to the axis, by rotation symmetry. That the
particle axis is the image of a circle in the plane $\hat{\bf
r}_{\rm e}=0$ of the wedge with center in the edge and passing
through the center of inversion, see the long dashes in Figs. 1
and 2, implies an eigenvector ${\bf u}^{(n)}$ of $\langle T
\rangle_{\rm wedge}^{(i,j)}$ tangent to this circle. Likewise,
degenerate eigenvectors perpendicular to the axis imply $d-1$
degenerate eigenvectors perpendicular to ${\bf u}^{(n)}$. The
simple form (\ref{stresswedge}) of the symmetric, traceless, and
conformal tensor density $\langle T \rangle$ for the wedge then
follows from scaling (dilatation symmetry), translation, and
reflection symmetry in the edge-subspace, and the
$\Omega$-independence of $\tau$ is due to the continuity equation
of the stress tensor, see Appendix A. In Appendix B we also
discuss, within the Gaussian model, the more complicated form of
the density $\langle T^{({\rm can})} \rangle_{\rm wedge}$ of the
{\it canonical} stress tensor, which also obeys the continuity
equation, but is not trace-free and not a conformal tensor, and
for which ${\bf u}^{(n)}(\Omega)$, in general, is not an
eigenvector. Its eigenvalues depend on both $\rho$ and $\Omega$,
and some of the eigenvalues diverge as the boundary planes of the
wedge are approached. The simple form (\ref{stresswedge}) is
recovered on adding the `improvement-term' \cite{bc80}.

We briefly comment on particle shapes where the {\it horizontal}
axis on the right hand side of Fig. 1 (passing through the small
triangle and square) is the rotation axis. These particles
resemble an apple (self-intersecting torus) for $\alpha < \pi$ and
an american football for $\alpha > \pi$ and could be conformally
mapped onto a {\it cone} with opening angle $\alpha$.

\section{GAUSSIAN MODEL WITH DIRICHLET BOUNDARIES}

In $d=2$ spatial dimensions, a wedge can be obtained from the half
plane by means of an appropriate conformal transformation. The
dependence on the angle $\alpha$ only enters via the
transformation and is to a large extent model-independent. For
example, for $i=j$, the stress-amplitude $\tau$ is independent of
the surface universality class $i$ and equals $[(\pi/\alpha)^2
-1]c/(24 \pi)$, where only the universal bulk-constant $c$, the
`conformal charge', depends on the model class. Using
(\ref{film}), this is consistent with the $i$-independent
film-amplitude $\Delta_{i,i}=-\pi c/24$ for a strip with equal
boundary conditions \cite{cardy}. Also the form of the scaling
functions $\bar{\cal P}_{i,i}={\cal A}_{\cal O}^{(i)} [(\alpha /
\pi) {\rm cos}(\pi \Omega / \alpha)]^{-x_{\cal O}}$ is to a large
extent model-independent \cite{be85} and is completely determined
by the bulk scaling index $x_{\cal O}$ and the universal, but
$i$-dependent, half-space boundary amplitude ${\cal A}_{\cal
O}^{(i)}$ of the scalar operator ${\cal O}$. For $i$ not equal to
$j$, the stress amplitude $\tau_{i,j}=-(c/(24
\pi))-\Delta_{i,j}/\alpha^2$ also depends on the boundary
universality classes. Since $\Delta_{i,j}$ for $i \not= j$ is in
general different from $-\pi c/24$, $\tau_{i,j}$ is nonvanishing
even for $\alpha=\pi$, i.e. for the half plane with a
non-homogeneous boundary \cite{buxu91}.

In $d>2$, however, no such conformal transformation exists, and
the $\alpha$-dependent quantities $\bar{\cal P}$ and $\tau$ depend
on the bulk and surface universality classes in a much stronger
way. In particular a bulk amplitude (like $c$) and the parallel
plate amplitude $\Delta_{i,j}$, are not sufficient to predict
$\tau_{i,j}(\alpha)$.

For the Gaussian Ginzburg-Landau field theory (\ref{Gaussham}) at
the critical point $t=0$ in $d>2$ spatial dimensions, the scaling
function ${\bar {\cal E}}$ of the energy density in the wedge is
given by
\be \label{phi2wedge}
{\bar {\cal E}} \, = \, - \langle \Phi^2 (\hat{\bf r})
\rangle_{\rm wedge} \, \rho^{d-2} \, / \, (\sqrt{2} \tilde{S}_d)
\quad , \quad \tilde{S}_d=\pi^{-d/2} \Gamma((d/2)-1)/4 \, ,
\ee
where we have suppressed the indices $(i,j)=({\rm D,D})$ with D
for Dirichlet \cite{edgenote}. The stress tensor in the Gaussian
model is the sum \cite{bc80}
\be \label{stressop}
T_{\kappa\lambda}(\hat{\bf r}) \, = \, T_{\kappa\lambda}^{\rm
(can)}(\hat{\bf r}) \, - \, I_{\kappa\lambda}(\hat{\bf r})
\ee
of the canonical stress tensor
\be \label{canonical}
T_{\kappa\lambda}^{\rm (can)}(\hat{\bf r}) \, = \,
(\hat{\partial}_{\kappa} \Phi(\hat{\bf r}))
(\hat{\partial}_{\lambda} \Phi(\hat{\bf r})) -
\delta_{\kappa\lambda}(\hat{\nabla}\Phi(\hat{\bf r}))^2 /2
\ee
and the so-called improvement term with
\be \label{improve}
I_{\kappa\lambda}(\hat{\bf r}) \, = \, \frac{1}{4} \frac{d-2}{d-1}
[\hat{\partial}_{\kappa} \hat{\partial}_{\lambda} -
\delta_{\kappa\lambda} \hat{\Delta}] \Phi^2 (\hat{\bf r}) \, .
\ee
Unlike $T^{({\rm can})}$, the density of $T$ in (\ref{stressop})
has a vanishing trace and transforms as a conformal tensor, see
Appendices B and C. The bulk amplitude $B_T$ in (\ref{stress2})
equals $[\tilde{S}_d (d-2)]^2 d/(d-1)$ for the Gaussian model.

The boundary-induced densities $\langle \Phi^2 (\hat{\bf r})
\rangle_{\rm wedge}$ and $\langle T_{\kappa \lambda} (\hat{\bf r})
\rangle_{\rm wedge}$ follow in an obvious way from the
boundary-induced contribution $\delta \langle
\hat{\varphi}_{12}\rangle_{\rm wedge}$ of the propagator in the
wedge
\begin{eqnarray} \label{propasum'}
\langle \hat{\varphi}_{12} \rangle_{\rm wedge} = \langle
\hat{\varphi}_{12} \rangle_{\rm bulk} + \delta \langle
\hat{\varphi}_{12} \rangle_{\rm wedge} \, ,
\end{eqnarray}
with
\be \label{phidach}
\hat{\varphi}_{12}=\Phi(\hat{\bf r}_1) \Phi(\hat{\bf r}_2) \, .
\ee
For the special opening angles $\alpha=\pi/g$, with $g$ a positive
integer, $\delta \langle \hat{\varphi}_{12} \rangle_{\rm wedge}$
is a linear combination of $2g-1$ bulk propagators from ${\bf
r}_1$ to images of ${\bf r}_2$, see Appendix B.

In order to determine $e_0$, $e_2$ in (\ref{smallomega}) and
$\tau$ in (\ref{stresswedge})  for {\it arbitrary} $\alpha$, we
use the representation of Ref. \onlinecite{cardywedge} at the
critical point
\begin{eqnarray} \label{propcardy}
\langle \hat{\varphi}_{12} \rangle_{\rm wedge} = \int d^{d-2} q \,
(2\pi)^{2-d} && e^{i{\bf q} (\hat{\bf r}_{e 1}-\hat{\bf r}_{e 2})}
\, (2/\alpha) \, \sum_{m=1}^{\infty} \, I_{m\pi/\alpha}(q
\rho_{<}) K_{m\pi/\alpha}(q \rho_{>}) \nonumber
\\
&&\times {\rm sin}\biggl[ m\pi\biggl( \frac{1}{2} +
\frac{\Omega_1}{\alpha} \biggr) \biggr] \, {\rm sin}\biggl[
m\pi\biggl( \frac{1}{2} + \frac{\Omega_2}{\alpha} \biggr) \biggr]
\, ,
\end{eqnarray}
where $\rho_{<}={\rm min}(\rho_1 , \rho_2)$, $\rho_{>}={\rm
max}(\rho_1 , \rho_2)$, and $I$ and $K$ are modified Bessel
functions. A more explicit expression arises for $(\hat{\bf
r}_{{\rm e} 1},\rho_1)=(\hat{\bf r}_{{\rm e} 2},\rho_2)$. Using
\begin{eqnarray} \label{rewrite}
\int_{0}^{\infty} \, d x \, x^{2A-1} \, I_{m\pi/\alpha}(x) \,
K_{m\pi/\alpha}(x) \, = \, 2^{2 A - 2} \,
\frac{\Gamma(A)}{\Gamma(1-A)} \, \int_{0}^{1} \, d t \, t^{m
\pi/\alpha} \, t^{A-1} \, (1-t)^{-2 A}
\end{eqnarray}
for $2A=d-2$ in order to rewrite the $q$-integral in a form where
the $m$-summation can be done \cite{cardywedge}, one finds
\begin{eqnarray} \label{propomega}
&& \tilde{S}_d^{-1}
\langle \hat{\varphi}_{12} \rangle_{\rm wedge}|_{(\hat{\bf
r}_{{\rm e} 1},\rho_1)=(\hat{\bf r}_{{\rm e} 2},\rho_2)} =
%
%
- (2/\alpha) \, {\rm sin}(\pi d/2)
\, \rho^{2-d} \,
\int_{0}^{1} \, d t \, \Psi_{d} (t) \, \nonumber \\
&& \, \times \Biggl\{ \frac{1-t^{\pi/\alpha}{\rm
cos}[(\Omega_1-\Omega_2)\pi/\alpha]}{1+t^{2\pi/\alpha}-2
t^{\pi/\alpha}{\rm cos}[(\Omega_1-\Omega_2)\pi/\alpha]} -
\frac{1+t^{\pi/\alpha}{\rm
cos}[(\Omega_1+\Omega_2)\pi/\alpha]}{1+t^{2\pi/\alpha}+2
t^{\pi/\alpha}{\rm cos}[(\Omega_1+\Omega_2)\pi/\alpha]}\Biggr\} \,
,
\end{eqnarray}
where $\rho \equiv \rho_1 = \rho_2$ and
\begin{eqnarray} \label{Psid}
\Psi_{d}(t) \, = \, t^{(d-4)/2} \, (1-t)^{2-d} \, .
\end{eqnarray}
The expression in Eq. (\ref{propomega}) satisfies the Dirichlet
boundary conditions since the curly bracket vanishes for
$\Omega_1$ (or $\Omega_2$) equal to $\alpha/2$ or $-\alpha/2$. For
$\Omega_1 \not= \Omega_2$ the $t$-integral converges in the
interval $2-(2\pi/\alpha)<d<4$ of spatial dimensions $d$ with the
lower and upper limits coming from the behaviors of the integrand
for $t \searrow 0$ and $t \nearrow 1$, respectively. The bulk
divergence $(\rho |\Omega_1-\Omega_2|)^{2-d}$ for
$|\Omega_1-\Omega_2| \to 0$ and $d>2$ contained in
(\ref{propomega}) comes from integrating the first term in curly
brackets over a region near $t=1$ where $1-t$ is of the order of
$|\Omega_1-\Omega_2|$.

A convenient way to evaluate the boundary-induced density
\begin{eqnarray} \label{limdelprop}
\langle \Phi^2(\hat{\bf r}) \rangle_{\rm wedge} = {\rm
lim}_{\hat{\bf r}_1 \to \hat{\bf r}, \hat{\bf r}_2 \to \hat{\bf
r}} \; \delta \langle \hat{\varphi}_{12} \rangle_{\rm wedge} \,
\end{eqnarray}
and its scaling function $\bar{\cal E}(\alpha,\Omega)$ in
(\ref{phi2wedge}) for $d=3$ by means of Eq. (\ref{propomega}), is
based on the observation \cite{examp} that $\bar{\cal E}$
is an {\it analytic} function of $d$ in an interval that includes
both $d=2$ and $d=3$ as interior points. Since the limit $\hat{\bf
r}_1 \to \hat{\bf r}, \hat{\bf r}_2 \to \hat{\bf r}$ of the bulk
propagator $\propto |\hat{\bf r}_1 - \hat{\bf r}_2|^{2-d}$ {\it
vanishes} for $d<2$ (while it is infinite for $d>2$), one may
replace $\delta \langle \hat{\varphi}_{12} \rangle_{\rm wedge}$ by
$\langle \hat{\varphi}_{12} \rangle_{\rm wedge}$ and use
(\ref{propomega}), in calculating the limit in (\ref{limdelprop})
for $d<2$. This leads to
\begin{eqnarray} \label{Ebar}
&&\bar{\cal E}(\alpha , \Omega) \, = \, (\sqrt{2}/\alpha) \, {\rm
sin} (\pi d/2) \; \int_0^1 \, d t \, \Psi_{d}(t) \nonumber \\
&& \qquad \qquad \times \Biggl\{ \frac{1}{1-t^{\pi/\alpha}} \, -
\, \frac{1+t^{\pi/\alpha} {\rm cos}(2\pi
\Omega/\alpha)}{1+t^{2\pi/\alpha}+2t^{\pi/\alpha}{\rm cos}(2\pi
\Omega/\alpha)} \Biggr\} \quad ; \quad d<2 \quad ,
\end{eqnarray}
in terms of an integral which is well defined for
$2-(2\pi/\alpha)<d<2$ and which has to be analytically continued
\cite{examp} in order to obtain $\bar{\cal E}$ for $d=3$.

For $e_0(\alpha)=\bar{\cal E}(\alpha , \Omega=0)$ the integral in
(\ref{Ebar}) becomes
\begin{eqnarray} \label{inte0}
J(\alpha , d) \, = \, \int_{0}^{1} \, d t \, \Psi_{d} (t) \,
t^{\pi/\alpha} \, \frac{2}{1-t^{2\pi/\alpha}} \, ,
\end{eqnarray}
and the continuation can be made by rewriting $J$ as the sum of
$J^{(1)}$ and $J^{(2)}$, where
\begin{eqnarray} \label{inte0'}
J^{(1)}(\alpha , d) \, = \, \int_{0}^{1} \, d t \, \Psi_{d} (t) \,
t^{\pi/\alpha} \, \Biggl[ \frac{2}{1-t^{2\pi/\alpha}} \, - \, l(t)
\Biggr] \, ,
\end{eqnarray}
and
\begin{eqnarray} \label{inte0''}
J^{(2)}(\alpha , d) \, = \, \int_{0}^{1} \, d t \, \Psi_{d} (t) \,
t^{\pi/\alpha} \, l(t) \, .
\end{eqnarray}
Here
\begin{eqnarray} \label{laue0}
l(t) \, = \, \frac{\alpha/\pi}{1-t} \, + \, 1 \, -
\frac{\alpha}{2\pi}
\end{eqnarray}
are the first two terms in the Laurent series of
$2/(1-t^{2\pi/\alpha})$ around $t=1$ so that the integrability
domain $2-(2\pi/\alpha)<d<4$ of $J^{(1)}$ extends up to $d=4$. As
a sum of beta-functions the continuation to $d=3$ of the integral
$J^{(2)}$ is trivial and yields $J^{(2)}(\alpha , 3)=-1$, implying
\begin{eqnarray} \label{e03}
e_0(\alpha)\, = \, (\sqrt{2} /\alpha) \, [1 \, - \, J^{(1)}(\alpha
, 3)] \quad , \quad d=3 .
\end{eqnarray}
Here $J^{(1)}(\alpha , 3)$ follows from the right hand side in Eq.
(\ref{inte0'}) on replacing $\Psi_{d}$ by $\Psi_{3}=t^{-1/2}
/(1-t)$, and for $\alpha$ arbitrary between $0$ and $2\pi$ we have
calculated it numerically. Using Eq. (\ref{isopart}), the
corresponding results for $a_1 B_{\epsilon}^{1/2} /L = {\rm
sin}(\alpha/2) e_0 (\alpha)$ in the interval $0<\alpha<\pi$ and
$a_1 B_{\epsilon}^{1/2} /{\cal D} = e_0 (\alpha)$ in the interval
$\pi<\alpha<2\pi$ are shown in Fig. 3. Analytical results for some
special values of $\alpha$ are given in Table 1.

In the Gaussian model some amplitudes of isotropic operators
beyond leading order in (\ref{small'}), such as $\Phi^4$ and
$\Phi^6$, are also determined \cite{bey} by $a_1$.

To calculate the small particle anisotropy-amplitude $b_1$ in
(\ref{aniso++part}), we need the coefficient $\propto \Omega^2$ of
$\bar{\cal E}$. The contribution of order $\Omega^2$ to the curly
bracket in (\ref{Ebar}) contains a factor $1-t^{\pi/\alpha}$ and
leads to a convergent integral up to $d=4$. Thus, no continuation
is necessary, and
\begin{eqnarray} \label{e23}
e_0(\alpha)\, e_2(\alpha) \, = \, (4 \sqrt{2} \pi^2 /\alpha^3) \,
\int_{0}^{1} \, d t \, \Psi_{3}(t) \, t^{\pi/\alpha} \, (1 -
t^{\pi /\alpha})/(1 + t^{\pi /\alpha})^3 \quad , \quad d=3 ,
\end{eqnarray}
which on using (\ref{aniso++part}) with (\ref{isopart}),
(\ref{e03}) leads to the results for $b_1$ shown in Fig. 4 and in
Table 1.

In Appendix D we use similar continuations in $d$ to calculate the
stress tensor amplitude $\tau(\alpha)$ from (\ref{stresswedge})
and (\ref{stressop})-(\ref{improve}) in $d=3$ with the result
(\ref{taugauss'}). The second anisotropy-amplitude $b_2$ then
follows from (\ref{b2}) and the value $d/B_T = 32 \pi^2 $, with
the results shown in Fig. 5 and Table 1.

It is interesting to compare dumbbells and lenses with ellipsoids.
We compare a dumbbell with $\alpha$ between $0$ and $\pi$ with a
prolate ellipsoid that circumscribes the dumbbell, touches it at
the highest and lowest points, and has the same curvature at these
points. Denoting by $D_{\parallel}$ and $D_{\perp}$ the diameters
of the ellipsoid parallel and perpendicular to the rotation axis,
\begin{eqnarray} \label{prolat}
D_{\parallel} \, = \, 2 L {\rm cos}^2(\alpha/4) \quad , \quad
D_{\perp} \, = \, \sqrt{2} L {\rm cos}(\alpha/4) \, .
\end{eqnarray}
Similarly, we compare a lens with $\alpha$ between $\pi$ and $2
\pi$ with an oblate circumscribing ellipsoid, so that
\begin{eqnarray} \label{oblat}
D_{\parallel} \, = \, {\cal D} {\rm ctg}(\alpha/4) \quad , \quad
D_{\perp} \, = \, {\cal D}
\end{eqnarray}
where ${\cal D} {\rm ctg}(\alpha/4)$ is the width \cite{eise04} of
the lens. The amplitudes $a_1, \, b_1,\, b_2$ of the ellipsoids
are shown as circles in Figs. 3-5. They follow from Refs.
\onlinecite{transl,fxi} and Eqs. (180)-(185) in Ref.
\onlinecite{PCI} where the long axis $[D_{\parallel} \, , \,
D_{\perp}]$ and short axis $[D_{\perp} \, , \, D_{\parallel}]$ of
the [prolate , oblate] ellipsoid is denoted by $l$ and $s$,
respectively.

As expected, the isotropic and anisotropic perturbations of the
polymer system from dumbbells are weaker and stronger,
respectively, than from the circumscribing prolate ellipsoids. The
oblate ellipsoids have stronger isotropic-perturbation amplitudes
$a_1$ and also slightly stronger anisotropic amplitudes $b_1$ and
$b_2$ than the lens.

\section{WEAK ANISOTROPY}

Consider the amplitudes $a_1, \, b_1$, and $b_2$ for particles
with a surface $S'$ which deviates only slightly from the surface
$S$ of a sphere with radius $R$. $S'$ is obtained by shifting each
surface point ${\bf r}_S$ of $S$ by a small amount $\eta
(\theta_S)$ toward the center of $S$ at the origin. Here
$\theta_S$ is the angle which ${\bf r}_S$ encloses with the
particle rotation axis, and we consider particles with a center of
reflection so that $\eta (\theta_S)=\eta (\pi - \theta_S)$. For a
dumbbell or lens with $\alpha = \pi + \delta \alpha$, we choose
$R=L/2$ and obtain
\begin{eqnarray} \label{etadumb}
\eta \, = \, \delta \alpha \, (L/4) \, |{\rm cos} \theta_S|
\end{eqnarray}
to first order in $\delta \alpha$. As expected from Fig. 1, $\eta$
in (\ref{etadumb}) is non-analytic at $\theta_S = \pi/2$. At the
end of Sec. III we have introduced apple- and football-shaped
particles with $\alpha = \pi + \delta  \alpha$ smaller and larger
than $\pi$, respectively. Nearly spherical particles of this
family are generated by
\begin{eqnarray} \label{etaapple}
\eta \, = \, \delta \alpha \, (L/4) \, {\rm sin} \theta_S \, .
\end{eqnarray}
For weakly anisotropic prolate and oblate ellipsoids with
$D_{\parallel} > D_{\perp}$ and $D_{\parallel} < D_{\perp}$,
respectively,
\begin{eqnarray} \label{etaellipsoid}
\eta \, = \, \frac{D_{\parallel} - D_{\perp}}{2} \, {\rm sin}^2
\theta_S \, ,
\end{eqnarray}
if we choose $R=D_{\parallel}/2$.

In the presence of the weakly anisotropic particle the propagator
is given by \cite{PCI}
\begin{eqnarray} \label{weakprop}
\langle {\varphi}_{12} \rangle \, = \, \langle {\varphi}_{12}
\rangle_{\rm sphere} \, + \, \int d S \, \eta (\theta_S) \,
\langle T_{\perp \, \perp} ({\bf r}_S) \, {\varphi}_{12}
\rangle_{\rm sphere}
\end{eqnarray}
to first order in $\eta$. Here $\int d S$ is an integral over the
surface $S$ of the sphere, and $T_{\perp \, \perp}$ is the
diagonal component of the stress tensor perpendicular to $S$. Due
to the Dirichlet boundary condition and the form
(\ref{stressop})-(\ref{improve}) of the stress tensor, $T_{\perp
\, \perp} ({\bf r}_S)$ can be replaced by
$(\partial_{\perp}\Phi)^2/2$ in the correlation function in
(\ref{weakprop}) with the result
\begin{eqnarray} \label{weakprop'}
\langle T_{\perp \, \perp} ({\bf r}_S) \, {\varphi}_{12}
\rangle_{\rm sphere} \, = \, [(d-2) \tilde{S}_d]^2 \,
\frac{(r_1^2-R^2)(r_2^2-R^2)}{R^2 (|r_1 - {\bf r}_S||r_2 - {\bf
r}_S|)^{d}} \, ,
\end{eqnarray}
and, from the behavior of (\ref{weakprop}) for $R \ll r_1, r_2$,
one finds \cite{weak}
\begin{eqnarray} \label{weaka1}
a_1 \sqrt{B_\epsilon} \, - \, R^{d-2}/\sqrt{2} \, = \, -
\frac{(d-2) \, \Gamma(d/2)}{\sqrt{2\pi} \, \Gamma((d-1)/2)} \,
R^{d-3} \, \int_0^\pi d \theta_S \, ({\rm sin}\theta_S)^{d-2} \,
\eta(\theta_S) \, ,
\end{eqnarray}
\begin{eqnarray} \label{weakb1}
b_1 \sqrt{B_\epsilon}/b_2 \, = \, - \frac{1}{4 \sqrt{2} \,
\pi^{d/2}} \frac{d+1}{d-1} \, \Gamma((d+2)/2) \, ,
\end{eqnarray}
and
\begin{eqnarray} \label{weakb2}
b_2 \, = \, R^{d-1} \, \frac{2 \pi^{(d-1)/2}}{\Gamma((d+1)/2)} \,
\int_0^\pi d\theta_S \, ({\rm sin}\theta_S)^{d-2} \, [d({\rm
cos}\theta_S)^2 - 1] \, \eta (\theta_S) \, .
\end{eqnarray}
Explicit expressions for dumbbells or lenses, apples or footballs,
and ellipsoids in an arbitrary spatial dimension $d$ follow on
inserting $\eta$ from (\ref{etadumb})-(\ref{etaellipsoid}). In
$d=3$,
\begin{eqnarray} \label{3weak}
&&a_1 \sqrt{B_\epsilon}/L \, = \, \frac{1}{2\sqrt{2}} \, - \,
\delta \alpha \biggl( \frac{1}{8\sqrt{2}} \, , \, \frac{\pi}{16
\sqrt{2}} \biggr) \, , \nonumber \\
&&b_2 / L^3 \, = \, \delta \alpha \biggl( \frac{\pi}{16} \, , \,
-\frac{\pi^2}{64} \biggr)
\end{eqnarray}
for (dumbbell or lens, apple or football), and
\begin{eqnarray} \label{3weak'}
&&a_1 \sqrt{B_\epsilon} \, = \, \frac{D_{\parallel}}{2\sqrt{2}} \,
+ \, \frac{D_{\perp}-D_{\parallel}}{3 \sqrt{2}} \, ,  \nonumber \\
&&b_2 \, = \, (D_{\perp}-D_{\parallel}) D^2 \, \frac{2 \pi}{15}
\end{eqnarray}
for ellipsoids. In our first order calculation $D$ may be either
$D_{\parallel}$ or $D_{\perp}$. Note that $b_2$ is negative for
the {\it prolate} shapes (dumbbell with $\delta \alpha <0$,
football with $\delta \alpha
>0$, and prolate ellipsoid with $D_{\parallel}>D_{\perp}$) and
positive for the {\it oblate} shapes (lens with $\delta \alpha
>0$, apple with $\delta \alpha <0$, and oblate ellipsoid with
$D_{\parallel}<D_{\perp}$).

The amplitudes $b_1$ follow from the amplitudes $b_2$ above via
the ratio
\begin{eqnarray} \label{3weakb1}
b_1 \sqrt{B_\epsilon}/b_2 \, = \, - \frac{3}{8 \pi \sqrt{2}} \, .
\end{eqnarray}

For a nearly spherical particle with rotation axis and reflection
symmetry, the ratio of the anisotropy amplitudes $b_1$ and $b_2$
is {\it independent} of its shape, see (\ref{weakb1}) and
(\ref{3weakb1}), but the ratio becomes shape-dependent for larger
deviations from spherical. For example,
\begin{eqnarray} \label{3strong}
b_1 \sqrt{B_\epsilon}/b_2 \, = \, - \frac{1}{\pi} \, \biggl(
\frac{7}{16 \sqrt{2}} \, , \, \frac{8+\sqrt{2}}{32} \, , \,
\frac{5}{16\sqrt{2}} \, , \, \frac{{\rm
ln}(2D_{\parallel}/D_{\perp})}{4\sqrt{2}} \biggr)
\end{eqnarray}
for a dumbbell of two touching spheres, a dumbbell with
$\alpha=\pi/2$, a disk, and an ellipsoidal needle with
$D_{\parallel} \gg D_{\perp}$. For the dumbbell-lens and ellipsoid
families the modulus of the negative ratio $b_1
\sqrt{B_\epsilon}/b_2$ is monotonically decreasing on increasing
$\alpha$ from $0$ to $2 \pi$ and $D_{\perp}/D_{\parallel}$ from
$0$ to $\infty$, respectively, i.e. on changing from more prolate
to more oblate shapes.

\section{INDUCED INTERACTIONS}

The expressions (\ref{inter}) for the free energy cost $F$ and
(\ref{small}) for the Boltzmann factor ${\rm exp}(-\delta {\cal
H})$ of a small particle determine the polymer-induced
orientation-dependent interactions. In particular, the leading
anisotropic interactions between a {\it particle and a wall},
\begin{eqnarray} \label{pwint}
F_{\rm aniso}^{\rm (p,w)} & = & p_0 \, ({\rm cos}^2 \vartheta_{\rm
P}) \, \{ b_1 \sqrt{B_{\epsilon}} 4 \pi \sqrt{2} {\cal M}_{\rm
h}'' (y)
\nonumber \\
&& \qquad - \, b_2 [f_0 (y) + (1/2) f_0 (y/2) ] \} \, ,
\end{eqnarray}
with the second derivative
\begin{eqnarray} \label{M''}
{\cal M}_{\rm h}'' \, = \, 4 [f_0 (y) - (1/2) f_0 (y/2)]
\end{eqnarray}
of the bulk-normalized polymer density in the half space
\cite{PCI} with respect to
\begin{eqnarray} \label{pwdist}
y  = z_{\rm P}/{\cal R}_g  \, , \, z_{\rm P}  = {\rm particle \,
wall \, distance} \, ,
\end{eqnarray}
and between {\it two particles} ${\rm P,Q}$,
\begin{eqnarray} \label{ABint}
&&F_{\rm aniso}^{\rm (P,Q)} \, = \, - p_0 \, ( {\rm cos}^2
\varphi_{\rm P} + {\rm cos}^2 \varphi_{\rm Q} ) \, \sqrt{2} \, a_1
\sqrt{B_{\epsilon}} \beta_{\rm VII} \, r_{\rm PQ}^{-1} \nonumber
\\
&& \qquad \qquad \times \{ 12 x^{-2} f_2 (x/2) + 6 x^{-1} f_1
(x/2) + f_0 (x/2) \} \, ,
\end{eqnarray}
with
\begin{eqnarray} \label{ABdist}
&&x  =  r_{\rm PQ}/{\cal R}_g  \, ,  \\
&&r_{\rm PQ}  = |{\bf r}_{\rm P} -{\bf r}_{\rm Q}|  = {\rm
particle \, particle \, distance} \,
\end{eqnarray}
and
\begin{eqnarray} \label{beta7}
\beta_{\rm VII}=[16 \sqrt{2} \pi b_1 \sqrt{B_{\epsilon}}+b_2]/2 \,
,
\end{eqnarray}
follow from (\ref{inter}) with the half space perturbed by ${\rm
exp}(-\delta {\cal H}) \propto \sigma_A ({\rm P})$ and the bulk
perturbed by ${\rm exp}(-\delta {\cal H}) \propto \sigma_I ({\rm
P}) \sigma_A ({\rm Q}) + \sigma_I ({\rm Q}) \sigma_A ({\rm P})$,
respectively. Here $f_n = {\rm i}^n {\rm erfc}$ is the n-fold
iterated complementary error function, $\vartheta_{\rm P}$ is the
angle between the particle axis and the surface normal of the
boundary wall, and $\varphi_{\rm P},\varphi_{\rm Q}$ are the
angles between the axes of particles ${\rm P,Q}$ and the distance
vector ${\bf r}_{\rm P} -{\bf r}_{\rm Q}$ of the two particles.
$F_{\rm aniso}^{\rm (P,Q)}$ is proportional to the anisotropic
part of $(\partial_{\parallel {\rm P}}^2 +
\partial_{\parallel {\rm Q}}^2) K(r_{\rm PQ})$, with $K$ the density-density
correlation function of ideal polymers in bulk solution.

For a dumbbell or lens with any $\alpha$ the particle-wall
expression predicts that, for small $y$, the particle orientation
{\it parallel} to the wall and, for large $y$, the {\it
perpendicular} orientation, have the lowest free energies. Note
that ${\cal M}_{\rm h}$ has a point of inflection, and ${\cal
M}_{\rm h}''$ is positive and negative for small and large $y$,
respectively. With the values $b_1$ and $b_2$ from Figures 4 and 5
above, the $b_1$ and $b_2$ contributions both favor the same,
parallel orientation for small $y$. For large $y$ they favor
different orientations, and their sum is $\propto - \beta_{\rm
VII} f_0(y/2)$, with $\beta_{\rm VII}$ from (\ref{beta7}) in which
$b_1$ dominates.

The two-particle expression predicts that particles align {\it
parallel} to their distance vector, as expected from the
attractive nature of the depletion interaction in a dilute polymer
solution.

Qualitatively similar behavior applies for prolate and oblate
ellipsoids, where $b_1$ and $b_2$ can be taken from Refs.
\onlinecite{ebm03,PCI,transl}.

\section{SUMMARY AND CONCLUDING REMARKS}

We have studied the interaction between long flexible nonadsorbing
polymers and mesoscopic colloidal dumbbells and lenses. The shape
of the colloids is characterized by a parameter $\alpha$, as shown
in Fig. 1 and ranges from two touching spheres for $\alpha=0$, to
a sphere for $\alpha=\pi$, and to a disk for $\alpha=2\pi$.

For small colloids and ideal polymers the amplitudes $a_1, \,
b_1$, and $b_2$ in the small particle expansion (\ref{small}),
which determine the isotropic and anisotropic features of the
interaction, are evaluated exactly for {\it arbitrary} $\alpha$.
They follow via the general relations (\ref{isopart}),
(\ref{aniso++part}), and (\ref{b2}) from the results (\ref{e03}),
(\ref{e23}), and (\ref{taugauss'}) for density profiles of the
Gaussian model in a wedge with opening angle $\alpha$ and
Dirichlet boundary conditions and are shown as crosses in Figs.
3-5. We compare with corresponding amplitudes for ellipsoids that
circumscribe and touch the dumbbells and lenses, see Eqs.
(\ref{prolat}), (\ref{oblat}). Their values are shown as circles
in Figs. 3-5. Analytical results for some special values of
$\alpha$ are collected in Table I.

We also consider weakly anisotropic particles of general shapes
with rotation axis and reflection center, see Eqs.
(\ref{weaka1})-(\ref{weakb2}). We find that the ratio $b_1/b_2$ in
Eqs. (\ref{weakb1}) and (\ref{3weakb1}) of the two
anisotropy-amplitudes of these particles is {\it independent} of
their shape. In particular we consider in Eqs. (\ref{3weak}) the
shapes of a self-intersecting torus which resembles an apple and
of an american football.

How to obtain from the amplitudes $a_1, \, b_1$, and $b_2$ the
orientation-dependent polymer-induced interaction between
particles is discussed in Sec. VI. While the preferential
alignment of two identical small particles is always parallel to
their distance vector, see Eq. (\ref{ABint}), the alignment of a
particle with respect to a wall changes from perpendicular to
parallel on decreasing the particle-wall distance, see Eq.
(\ref{pwint}). It would be interesting to check our predictions
with simulations or real experiments.

The simple and general forms (\ref{stresswedge}) and
(\ref{stresspart}) of the density of the conformal stress tensor
in a wedge and outside a dumbbell or lens, with eigenvectors shown
in Fig. 2, follow from combining symmetries of the two geometries,
see the end of Sec. III. We show in Appendix C that the sum
(\ref{stressop}) of canonical tensor and improvement term is a
conformal tensor, while the separate terms are not and have
densities with a more complicated form as discussed in Appendix B.

\begin{acknowledgments}
We thank T.W. Burkhardt for useful discussions.
\end{acknowledgments}
%


\appendix

\section{Continuity equation in the wedge}

Here we show that the $\Omega$-independence of the prefactor
$\tau_{i,j}$ in (\ref{stresswedge}) follows from the continuity
equation. For convenience we choose Cartesian axes perpendicular
to the edge in the $\Omega=0$ and $\Omega=\pi/2$ half planes and
denote them by indices $v$ and $w$, respectively, so that
\begin{eqnarray} \label{uvw}
&&(\hat{r}_v \, , \, \hat{r}_w) \, = \, \rho \, ({\rm cos}\Omega
\, , \, {\rm sin}\Omega) \, , \nonumber \\
&&(u_{v}^{(n)} \, , \, u_{w}^{(n)}) \, = \, (-{\rm sin}\Omega \, ,
\, {\rm cos}\Omega) \, ,
\end{eqnarray}
and
\begin{eqnarray} \label{chainrule}
\left( \begin{array}{ccc}
\rho \partial \Omega/\partial \hat{r}_v &,& \rho \partial \Omega/\partial \hat{r}_w \\
\partial \rho/\partial \hat{r}_v &,& \partial \rho/\partial \hat{r}_w \end{array}
\right) \,\, = \,
\left( \begin{array}{ccc}
-{\rm sin}\Omega &,& {\rm cos}\Omega \\
{\rm cos}\Omega &,& {\rm sin}\Omega \end{array} \right) \, .
\end{eqnarray}
Substituting (\ref{stresswedge}) with $\tau_{i,j}(\alpha) \to
\tau_{i,j}(\alpha,\Omega)$ into the two continuity equations
\cite{cardy}
\begin{eqnarray} \label{conserv}
&&\partial \langle T_{vv} \rangle_{\rm wedge}/ \partial
\hat{r}_{v} \, +
\, \partial \langle T_{vw} \rangle_{\rm wedge}/ \partial \hat{r}_{w} \, = \, 0 \nonumber \\
&&\partial \langle T_{wv} \rangle_{\rm wedge}/ \partial
\hat{r}_{v} \, + \, \partial \langle T_{ww} \rangle_{\rm wedge}/
\partial \hat{r}_{w} \, = \, 0 \, ,
\end{eqnarray}
using (\ref{uvw}), and calculating the derivatives by means of the
chain rule and (\ref{chainrule}), one finds that the
$\rho$-derivatives of the prefactor $\rho^{-d}$ cancel the
$\Omega$-derivatives of the eigenvector ${\bf u}^{(n)}$. Only the
$\Omega$-derivatives of $\tau$ remain, and the left hand sides of
the first and second Eq. (\ref{conserv}) are given by $(\partial
\tau / \partial \Omega) (d-1)/\rho^{d+1}$ multiplied by ${\rm
sin}\Omega$ and $-{\rm cos}\Omega$, respectively. Thus each of the
two equations implies that $\tau$ is independent of $\Omega$.

\section{Wedge with ${\bf \alpha=\pi/2}$}

The propagator $\langle \hat{\varphi}_{12} \rangle_{\rm wedge}$
for the Gaussian model in a wedge with Dirichlet boundary
conditions and $\alpha=\pi/g$, with $g$ an integer, can be
obtained by the method of images. Besides the half space
($\alpha=\pi$) we consider here the simplest case of a wedge with
$\alpha=\pi/2$ in which there are three images. With the notation
\be \label{short}
\hat{r}_{x i} \, = \, X_i \; , \; \hat{r}_{y i} \, = \, Y_i
\ee
for the components of
\be \label{xy}
\hat{\bf r}_i \, = \, (\hat{\bf r}_{{\rm e} i}, \, \hat{r}_{x i},
\, \hat{r}_{y i}) \quad ; \quad i=1,2
\ee
perpendicular to the edge along Cartesian axes in the two boundary
half planes $\Omega=-\pi/4$ (i.e. $Y=0, X \ge 0$) and
$\Omega=\pi/4$ (i.e. $X=0, Y \ge 0$), the propagator at the
critical point has the form (\ref{propasum'}) with
\be \label{bulkexy}
\langle \hat{\varphi}_{12} \rangle_{\rm bulk} \, = \, \tilde{S}_d
\, [{\bf e}_{12}^2 + (X_1 - X_2)^2 + (Y_1 - Y_2)^2]^{-(d-2)/2} \,
,
\ee
and
\begin{eqnarray} \label{propa}
\delta \langle \hat{\varphi}_{12} \rangle_{\rm wedge} \, = \,
\tilde{S}_d \, \{&-[{\bf e}_{12}^2 + (X_1 + X_2)^2 + (Y_1 -
Y_2)^2]^{-(d-2)/2}& \nonumber
\\
&+ [{\bf e}_{12}^2 + (X_1 + X_2)^2 +
(Y_1 + Y_2)^2]^{-(d-2)/2}& \nonumber \\
&-[{\bf e}_{12}^2 + (X_1 - X_2)^2 + (Y_1 + Y_2)^2]^{-(d-2)/2}& \}
\, .
\end{eqnarray}
Here ${\bf e}_{12}^2 \equiv (\hat{\bf r}_{{\rm e} 1}-\hat{\bf
r}_{{\rm e} 2})^2$, and $\tilde{S}_d$ is defined in
(\ref{phi2wedge}).

Letting $\hat{\bf r}_1 \to \hat{\bf r}, \, \hat{\bf r}_2 \to
\hat{\bf r}$ on the right hand side of (\ref{propa}) leads to the
boundary-induced profile
\be \label{engypi2}
-\bar{\cal E} \, (\rho/2)^{2-d} \sqrt{2} \, \equiv \, \langle
\Phi^2(\hat{\bf r}) \rangle_{\rm wedge} 2^{d-2}/\tilde{S}_d \, =
\, - X^{2-d} \, - \, Y^{2-d} \, + \, \rho^{2-d}
\ee
of the energy density in (\ref{phi2wedge}), which  due to
\be \label{XYOmro}
X \equiv \hat{r}_x = \rho ({\rm cos}\Omega - {\rm
sin}\Omega)/\sqrt{2} \; , \; Y \equiv \hat{r}_y = \rho ({\rm
cos}\Omega + {\rm sin}\Omega)/\sqrt{2}
\ee
is in accordance with the general form (\ref{scalwedge}) of a
scalar density with bulk exponent $x_{\cal O}=d-2$. It diverges on
approaching the boundary planes $\Omega = \pm \pi/4$.

For the boundary-induced densities of the canonical stress tensor
(\ref{canonical}) and the `improvement'-term (\ref{improve}),
\be \label{shortstress}
\bigl( \langle T_{\kappa\lambda}^{\rm (can)}(\hat{\bf r})
\rangle_{\rm wedge} \; , \; \langle - I_{\kappa\lambda}(\hat{\bf
r}) \rangle_{\rm wedge} \bigr) \, = \, (d-2) \tilde{S}_d 2^{-d} \,
\bigl( {\cal T}_{\kappa\lambda} \, , \, -{\cal J}_{\kappa\lambda}
\bigr) \, ,
\ee
Eq. (\ref{propa}) yields
\begin{eqnarray} \label{Tcpi2}
&&{\cal T}_{xx} \, = \, (d-2) Y^{-d} + d \rho^{-d-2} X^2 -
(d-1) \rho^{-d} \nonumber \\
&&{\cal T}_{xy} \, = \, d \rho^{-d-2} X Y \nonumber \\
&&{\cal T}_{{\rm e}\beta {\rm e}\beta} \, = \, (d-2) \bigl( X^{-d}
+ Y^{-d} \bigr) - (d-3) \rho^{-d} \,
\end{eqnarray}
and
%
\begin{eqnarray} \label{Ipi2}
&&-{\cal J}_{xx} \, = \, -(d-2) Y^{-d} - [d(d-2)/(d-1)]
\rho^{-d-2} X^2 +
(d-2) \rho^{-d} \nonumber \\
&&-{\cal J}_{xy} \, = \, -[d(d-2)/(d-1)] \rho^{-d-2} X Y \nonumber \\
&&-{\cal J}_{{\rm e}\beta {\rm e}\beta} \, = \, -(d-2) \bigl(
X^{-d} + Y^{-d} \bigr) + [(d-2)^2 /(d-1)] \rho^{-d} \, .
\end{eqnarray}
Here ${\rm e}\beta$ with $\beta=3,..,d$ runs over the $d-2$
Cartesian directions of the edge-subspace. ${\cal T}_{yy}$ and
$-{\cal J}_{yy}$ follow from the above expressions for ${\cal
T}_{xx}$ and $-{\cal J}_{xx}$ on exchanging $X$ and $Y$. All other
components of $\langle T^{\rm (can)} \rangle$ and $\langle -I
\rangle$ vanish by symmetry.

In the sum ${\cal T}_{\kappa\lambda}-{\cal J}_{\kappa\lambda}$ all
contributions proportional to $X^{-d}$ and $Y^{-d}$ cancel, and
the density of the stress tensor $T$ in (\ref{stressop}) has the
simple form of (\ref{stresswedge}) with
\be \label{taupi2}
\tau (\alpha=\pi/2) \, = \, 2^{-d} \, \frac{d-2}{d-1} \,
\tilde{S}_d \,
\ee
implying
\be \label{b2pi2}
b_2 / L^d \, = \, -2(\pi/8)^{d/2} / \Gamma(d/2)
\ee
if one uses (\ref{b2}) with ${\cal D}/L=1/\sqrt{2}$ and the forms
of $B_T$ below (\ref{improve}) and $\tilde{S}_d$ in
(\ref{phi2wedge}).

However, the densities of $T^{\rm (can)}$ and $I$ separately have
nonvanishing traces, with
\begin{eqnarray} \label{tracecan}
{\cal T}_{xx}+{\cal T}_{yy}+\sum_{\beta=3}^d {\cal T}_{{\rm
e}\beta {\rm e}\beta} &=& - \langle (\nabla \Phi)^2 \rangle
2^{d-1} / \tilde{S}_d \nonumber \\
&=& (d-2)[(d-1)(X^{-d} + Y^{-d}) - (d-2) \rho^{-d}] \, ,
\end{eqnarray}
and a more complicated form. For example, their diagonal
components parallel to the edge,  and the trace (\ref{tracecan}),
depend  not only on $\rho$ but, due to the terms $\propto X^{-d} +
Y^{-d}$, also on $\Omega$, and diverge on approaching the
boundaries $X=0$ or $Y=0$. There is a corresponding divergence of
the diagonal components of $T^{\rm (can)}$ and $I$ parallel to a
Dirichlet boundary wall for the half space ($\alpha=\pi$).
Moreover, ${\bf u}^{(n)}$ of Eq. (\ref{stresswedge}) and Fig. 2
is, in general, not an eigenvector of $\langle T^{\rm (can)}
\rangle$, due to the terms proportional to $Y^{-d}$ and $X^{-d}$
in ${\cal T}_{xx}$ and ${\cal T}_{yy}$, respectively. In the
continuity equation satisfied by $\langle T^{\rm (can)} \rangle$,
these terms do not contribute.

\section{Conformal stress density}

Here we verify that in the Gaussian model the boundary-induced
profile $\langle T \rangle$, with $T$ the sum (\ref{stressop}) of
the canonical stress tensor and the improvement term, is a
conformal tensor. For a conformal transformation $\hat{\bf r}' \to
{\bf r}'$ which leads from a geometry $\hat{G}'$ to a geometry
$G'$, we use the transformation law
\be \label{proptrafo}
\delta \langle \Phi ({\bf r}_{1}') \, \Phi ({\bf r}_{2}')
\rangle_{G'} \, = \, (b(\hat{\bf r}_{1}') \, b(\hat{\bf
r}_{2}'))^{(d-2)/2} \, \delta \langle \Phi (\hat{\bf r}_{1}') \,
\Phi (\hat{\bf r}_{2}') \rangle_{\hat{G}'}
\ee
for the propagator $\delta \langle\Phi\Phi \rangle$ with bulk
contribution subtracted as in (\ref{propasum'}). The dilatation
factor $b(\hat{\bf r}')$ of the transformation is defined below
Eq. (\ref{Cdistance}). Consider the boundary-induced density
$\langle T_{kl}({\bf r}') \rangle_{G'}$, where $T_{kl}$ follows
from (\ref{stressop}) on replacing $(\kappa,\lambda,\hat{\bf r})$
by $(k,l,{\bf r}')$. As in Appendix B each of its contributions
follows from the subtracted propagator on the left hand side of
(\ref{proptrafo}) by appropriate differentiations, where for the
terms in $-\langle I_{kl} \rangle_{G'}$ and $\langle T_{kl}^{(\rm
can)} \rangle_{G'}$ the arguments ${\bf r}_{1}'$ and ${\bf
r}_{2}'$ are set equal to ${\bf r}'$ before and after
differentiating, respectively. Following the same steps on the
right hand side of (\ref{proptrafo}), one finds that in the sum
$\langle T_{kl} \rangle_{G'} = \langle T_{kl}^{(\rm can )}
\rangle_{G'} - \langle I_{kl} \rangle_{G'}$ the contributions
involving ${\bf r}'$-derivatives of $b$ that come from calculating
$T^{(\rm can)}$ {\it cancel} the contributions involving ${\bf
r}'$-derivatives of $\partial \hat{\bf r}' /
\partial {\bf r}'$ and $b$ that come from calculating $-I$
and one is left with the transformation formula
\be \label{stresstrafo}
\langle T_{kl} ({\bf r}') \rangle_{G'} \, = \, \sum_{\kappa ,
\lambda} \, b^{d-2} (\hat{\bf r}') \, \Biggl( \frac{\partial
\hat{r}_{\kappa}'}{\partial r_{k}'} \Biggr) \, \Biggl(
\frac{\partial \hat{r}_{\lambda}'}{\partial r_{l}'} \Biggr) \,
\langle T_{\kappa \lambda} (\hat{\bf r}') \rangle_{\hat{G}'} \,
\ee
of a conformal tensor. In general $T^{({\rm can})}$ and $I$
separately do not satisfy a transformation law of the form of
(\ref{stresstrafo}). It is instructive to explicitly check the
above arguments for the inversion $\hat{\bf r}' = L^2 {\bf r}' /
r'^2$, where
\be \label{invmatrix}
\frac{\partial \hat{r}_{\kappa}'}{\partial r_{k}'} \, = \,
b(\hat{\bf r}') \, {\cal I}_{\kappa , k}({\bf r}') \quad , \quad
b(\hat{\bf r}') \, = \, \frac{L^2}{r'^2} \, ,
\ee
with ${\cal I}$ defined in Eq. (\ref{I}), and where
(\ref{stresstrafo}) reduces to the transformation law of Eqs.
(2.36) and (2.37) in Ref. \onlinecite{eise04}.

The transformation formula (\ref{stresstrafo}) for the stress
tensor density is also expected to apply beyond the Gaussian
model. Given in geometry $\hat{G}'$ an eigenvector of $\langle T
(\hat{\bf r}') \rangle_{\hat{G}'}$ which points along a certain
distance vector $d \hat{\bf r}'$ and belongs to an eigenvalue
$\hat{t}$, Eq. (\ref{stresstrafo}) implies in geometry ${G}'$ an
eigenvector of $\langle T ({\bf r}') \rangle_{G'}$ which points
along the corresponding distance vector $d {\bf r}'$ and belongs
to the eigenvalue $b^{d} (\hat{\bf r}') \hat{t}$.

\section{Stress amplitude for arbitrary opening angle}

It is convenient to calculate $\tau(\alpha)$ from
(\ref{stresswedge}) with $\kappa=\lambda$ equal to an edge
direction ${\rm e} \beta$. Using (\ref{stressop})-(\ref{improve})
and the property
\be \label{vantrace}
2 \, \langle (\nabla \Phi)^2 \rangle_{\rm wedge} \, = \, \Delta \,
\langle \Phi^2 \rangle_{\rm wedge}
\ee
of the vanishing trace of $\langle T^{\rm (can)}-I \rangle_{\rm
wedge}$, this yields
\be \label{taugauss}
\tau (\alpha) \, = \,  \rho^d \langle T_{{\rm e} \beta {\rm e}
\beta} \rangle \,= \, \rho^d  \biggl[\langle (\Phi_{{\rm e}
\beta})^2 \rangle_{\rm wedge} - \frac{1}{4(d-1)} \Delta \langle
\Phi^2 \rangle_{\rm wedge} \biggr] \, ,
\ee
where
\be \label{Phiebeta}
\Phi_{{\rm e} \beta}(\hat{\bf r}) \, = \, \partial \Phi (\hat{\bf
r}) / \partial \hat{r}_{{\rm e} \beta} \, .
\ee
Since the boundary induced profile
\be \label{limPhiebeta}
\langle (\Phi_{{\rm e} \beta}(\hat{\bf r}))^2 \rangle_{\rm wedge}
= {\rm lim}_{\hat{\bf r}_1 \to \hat{\bf r} , \hat{\bf r}_2 \to
\hat{\bf r}} [\langle \Phi_{{\rm e} \beta} (\hat{\bf r}_1)
\Phi_{{\rm e} \beta} (\hat{\bf r}_2) \rangle_{\rm wedge}-\langle
\Phi_{{\rm e} \beta} (\hat{\bf r}_1) \Phi_{{\rm e} \beta}
(\hat{\bf r}_2) \rangle_{\rm bulk}]
\ee
is, apart from a factor $\tilde{S}_d$, analytic \cite{examp} in
$d$, the calculation proceeds similar to
(\ref{limdelprop})-(\ref{e03}): We consider $d<0$, where
\be \label{bulkPhiebeta}
\langle \Phi_{{\rm e} \beta} (\hat{\bf r}_1) \Phi_{{\rm e} \beta}
(\hat{\bf r}_2) \rangle_{\rm bulk}=(d-2) [|\hat{\bf r}_1-\hat{\bf
r}_2|^2 - d (\hat{r}_{1{\rm e} \beta}-\hat{r}_{2{\rm e} \beta})^2
] |\hat{\bf r}_1-\hat{\bf r}_2|^{-2-d} \, \tilde{S}_d
\ee
does not contribute on the right hand side of (\ref{limPhiebeta}),
and find for $\Omega=0$
\be \label{Phiebeta2}
\rho^d \, \langle (\Phi_{{\rm e} \beta}(\hat{\bf r}))^2
\rangle_{\rm wedge} |_{\Omega=0} \,  = \, 2 \pi^{(2-d)/2}
\frac{1}{\alpha \Gamma (1-(d/2))} \, J_{\rm e} \, ,
\ee
with
\be \label{Phiebeta3}
J_{\rm e} \, = \, \int_0^1 d t \, \Xi_d (t) \, t^{\pi/\alpha}
\frac{1}{1-t^{2\pi/\alpha}} \, ,
\ee
where
\be \label{Xi}
\Xi_d (t) \, = \, t^{(d-2)/2} (1-t)^{-d} \, .
\ee
Here we have rewritten $\langle (\Phi_{{\rm e} \beta})^2
\rangle_{\rm wedge}$ as $\sum_{\beta} \langle (\Phi_{{\rm e}
\beta})^2 \rangle_{\rm wedge} /(d-2)$ and used Eqs.
(\ref{propcardy}) and (\ref{rewrite}) with $2A=d$. The integral
(\ref{Phiebeta3}) converges for $- 2 \pi /\alpha < d < 0$.

Writing $J_{\rm e}$ as the sum of
\be \label{Je1}
J_{\rm e}^{(1)}=\int_0^1 d t \, \Xi_d (t) \, t^{\pi/\alpha} \,
\biggl[ \frac{1}{1-t^{2\pi/\alpha}}-k(t)\biggr]
\ee
and
\be \label{Je2}
J_{\rm e}^{(2)}=\int_0^1 d t \, \Xi_d (t) \, t^{\pi/\alpha} \,
k(t) \, ,
\ee
with
\be \label{kt}
k(t)= \frac{\alpha}{2\pi} \frac{1}{1-t}+\frac{1}{2} \biggl( 1-
\frac{\alpha}{2 \pi} \biggr) + \frac{1}{24} \biggl( \frac{2 \pi
}{\alpha}-\frac{\alpha}{2 \pi} \biggr) [2(1-t)+(1-t)^2] \, ,
\ee
the continuation to $d=3$ leads to
\be \label{Je2in3}
J_{\rm e}^{(2)} (\alpha,3) \, = \, \frac{1}{72} \, \biggl[ \biggl(
\frac{2 \pi}{\alpha} \biggr)^2 -1 + 6 \frac{\alpha}{2 \pi} \biggr]
\, ,
\ee
yielding
\be \label{Phiein3}
\langle (\Phi_{{\rm e} \beta}(\hat{\bf r}))^2 \rangle_{\rm wedge}
|_{\Omega=0} \, \rho^d \, = \, - \frac{1}{\alpha \pi} [J_{\rm
e}^{(1)}(\alpha,3)+J_{\rm e}^{(2)}(\alpha,3)]
\ee
in $d=3$ where $J_{\rm e}^{(1)} (\alpha,3)$ follows from
(\ref{Je1}) on replacing $\Xi_d$ by $\Xi_3=t^{1/2}(1-t)^{-3}$.

In the second contribution to $\tau$ in (\ref{taugauss}),
\be \label{DeltaPhi2}
\rho^d \, \Delta \langle \Phi^2 \rangle_{\rm wedge} \, = \, -
\sqrt{2} \tilde{S}_d \{ \partial_{\Omega}^2 \bar{\cal E} + (d-2)^2
\bar{\cal E} \} \,
\ee
can, for $\Omega=0$ and $d=3$, be expressed by $e_0$ and $e_2$ in
(\ref{e03}), (\ref{e23}), so that
\be \label{taugauss'}
\tau (\alpha) \, = \, - \frac{1}{\alpha \pi} \, [J_{\rm
e}^{(1)}(\alpha,3)+J_{\rm e}^{(2)}(\alpha,3)] \, + \, \frac{1}{16
\sqrt{2} \pi} e_0 (\alpha) [e_2 (\alpha) + 1] \, .
\ee
\newpage
\newpage
\begin{table}
\begin{center}
\begin{tabular}{c||c|c|c|||c||c}
   \hspace*{0.7cm}$\alpha$\hspace*{0.7cm}&\hspace*{1.1cm}0\hspace*{1.1cm}&
   \hspace*{0.8cm}$\pi/2$\hspace*{0.8cm}&\hspace*{1.0cm}$\pi$\hspace*{1.0cm}&
   \hspace*{1.0cm}$2\pi$\hspace*{1.0cm}&\hspace*{0.7cm}$\alpha$\hspace*{0.7cm} \\ \hline
   & $2^{-1/2} {\rm ln}2$ & $2^{-1/2}-2^{-2}$ & $2^{-3/2}$ & $2^{-1/2} \pi^{-1}$ &   \\
 $a_1 \sqrt{B_{\epsilon}} /L$    & $=0.490$ &  $=0.457$  &  $=0.354$  &  $=0.225$ &   $a_1 \sqrt{B_{\epsilon}} /{\cal D}$   \\
 & $(0.569)$ & $(0.508)$  &   &  &   \\  \hline
 & $7 \zeta (3) /(64 \sqrt{2})$ & $(4+2^{-1/2})/(64 \sqrt{2})$ & & $-5/(96 \pi \sqrt{2})$ &   \\
 $b_1 \sqrt{B_{\epsilon}} /L^3$  & $=0.0930$ & $=0.0520$  & 0 & $=-0.0117$  & $b_1 \sqrt{B_{\epsilon}} /{\cal D}^3$     \\
 & $(0.0561)$ & $(0.0304)$ &  &  &   \\ \hline
 & $-\pi \zeta (3) /4$& $-\pi 2^{-5/2}$  &   &  $1/6$  &   \\
 $b_2  /L^3$ & $=-0.944$ & $=-0.555$  &  0  &  $=0.166$  & $b_2  /{\cal D}^3$ \\
 & $(-0.630)$ & $(-0.346)$ &  &   &

\end{tabular}
%
%
\end{center}
\caption{Amplitudes for dumbbells with $\alpha=0$, $\pi/2$, the
sphere ($\alpha=\pi$), and the disk ($\alpha=2\pi$) in three
spatial dimensions. Numbers in brackets denote amplitudes for
prolate ellipsoids circumscribing the dumbbells. The values
correspond to the crosses and circles in Figs. 3-5.}
\label{smallparts}
\end{table}
%
%
%

%
\newpage
\begin{figure}[thb]
\includegraphics*[width=0.7\textwidth]{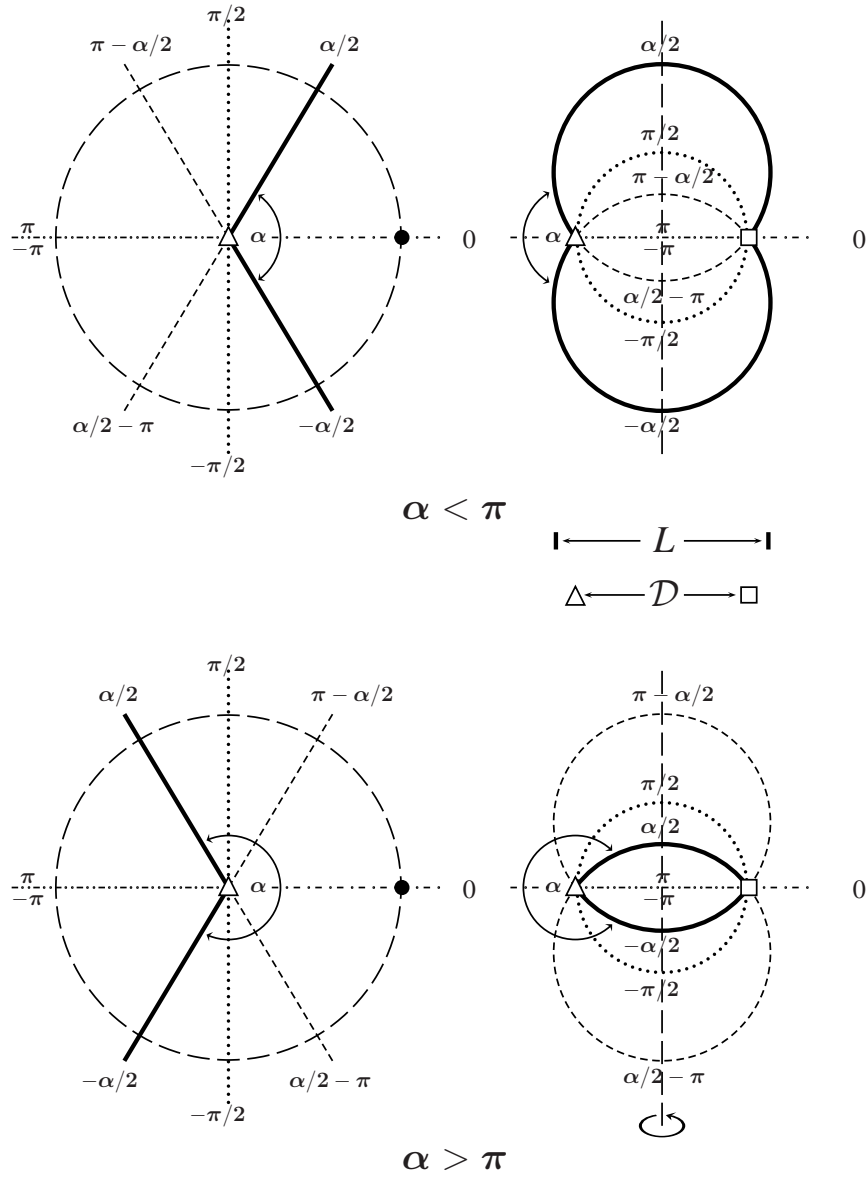}
%
%
\caption{\label{fig1} Conformal mapping of a wedge onto a dumbbell
or lens.}
\end{figure}
\newpage
\begin{figure}[thb]
\includegraphics*[width=0.55\textwidth,angle=90]{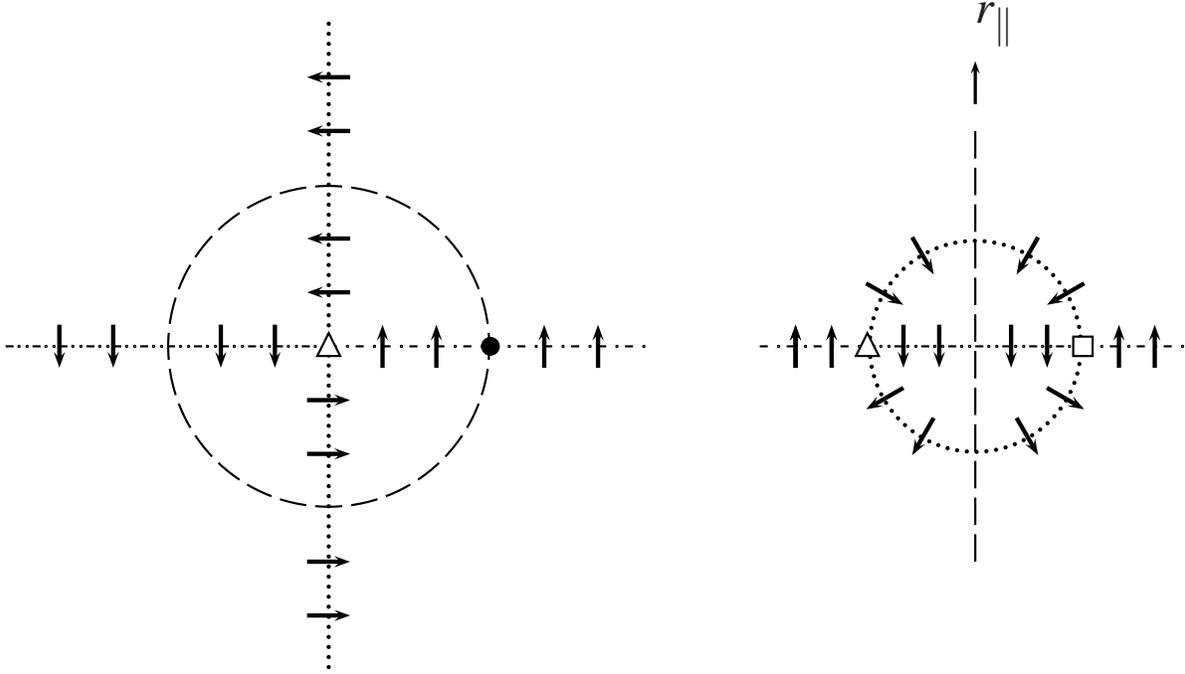}
%
%
\caption{\label{fig2} The normalized eigenvectors ${\bf
u}^{(n)}(\Omega)$ (left) and ${\bf u}^{(N)}({\bf r})$ (right) of
the stress tensor density in a wedge and outside a dumbbell or
lens. The vector field ${\bf u}^{(N)}$ is determined by Eq.
(\ref{uN}). The symbols for lines and points correspond to those
in Fig. 1.}
\end{figure}
\newpage
\begin{figure}[thb]
\includegraphics*[width=0.7\textwidth,angle=90]{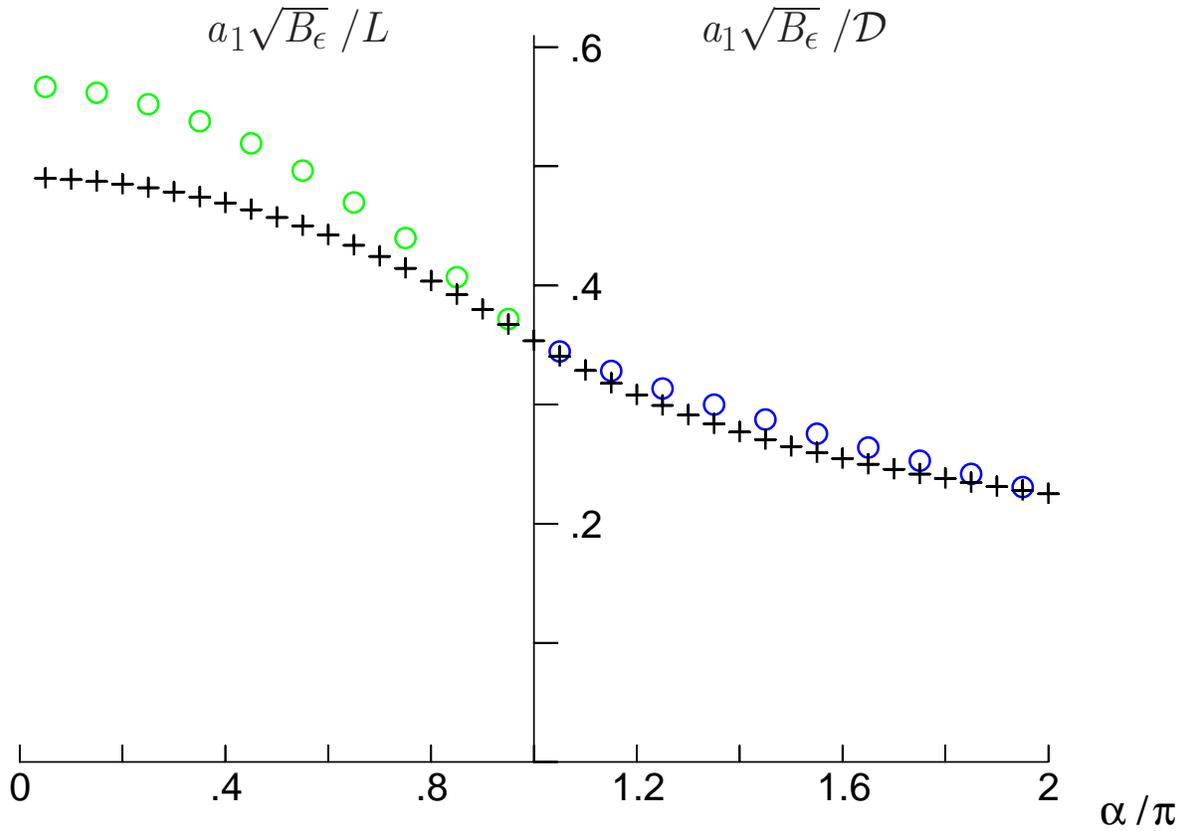}
%
%
\caption{\label{fig3} The leading amplitude $a_1$ in (\ref{small})
for particles with Dirichlet boundary conditions in the Gaussian
model. Crosses denote results for dumbbells ($0 < \alpha/\pi < 1$)
and lenses ($1 < \alpha/\pi < 2$). Circles denote results for
circumscribing prolate and oblate ellipsoids, see the paragraph
containing Eqs. (\ref{prolat}) and (\ref{oblat}). The oblate
ellipsoid and the lens coincide for $\alpha/\pi=1$ and 2, where
they become a sphere and a disk, respectively.}
\end{figure}
\newpage
\begin{figure}[thb]
\includegraphics*[width=0.7\textwidth,angle=90]{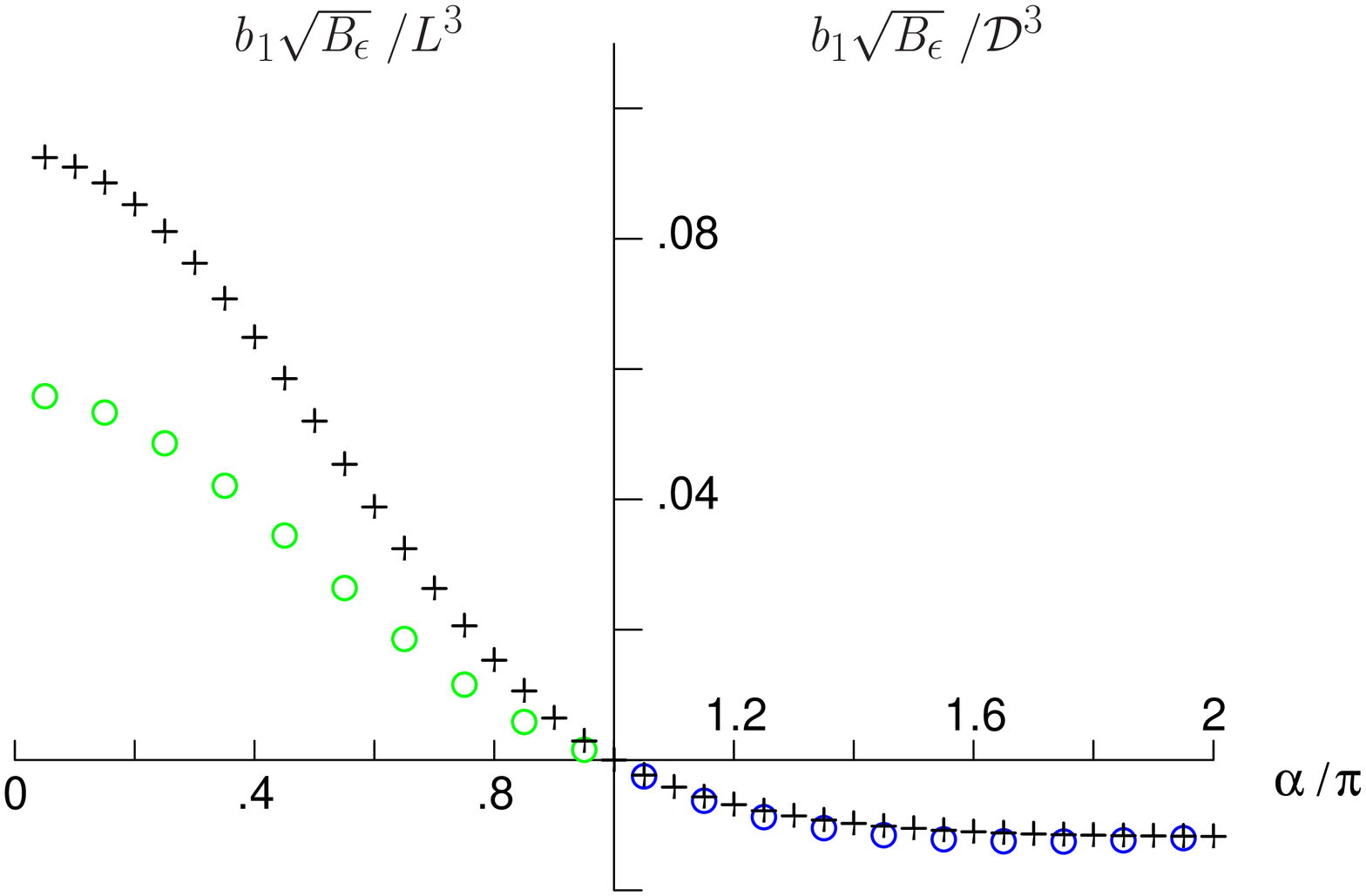}
%
%
\caption{\label{fig4} The anisotropy amplitude $b_1$ in
(\ref{small}).}
\end{figure}
\newpage
\begin{figure}[thb]
\includegraphics*[width=0.7\textwidth,angle=90]{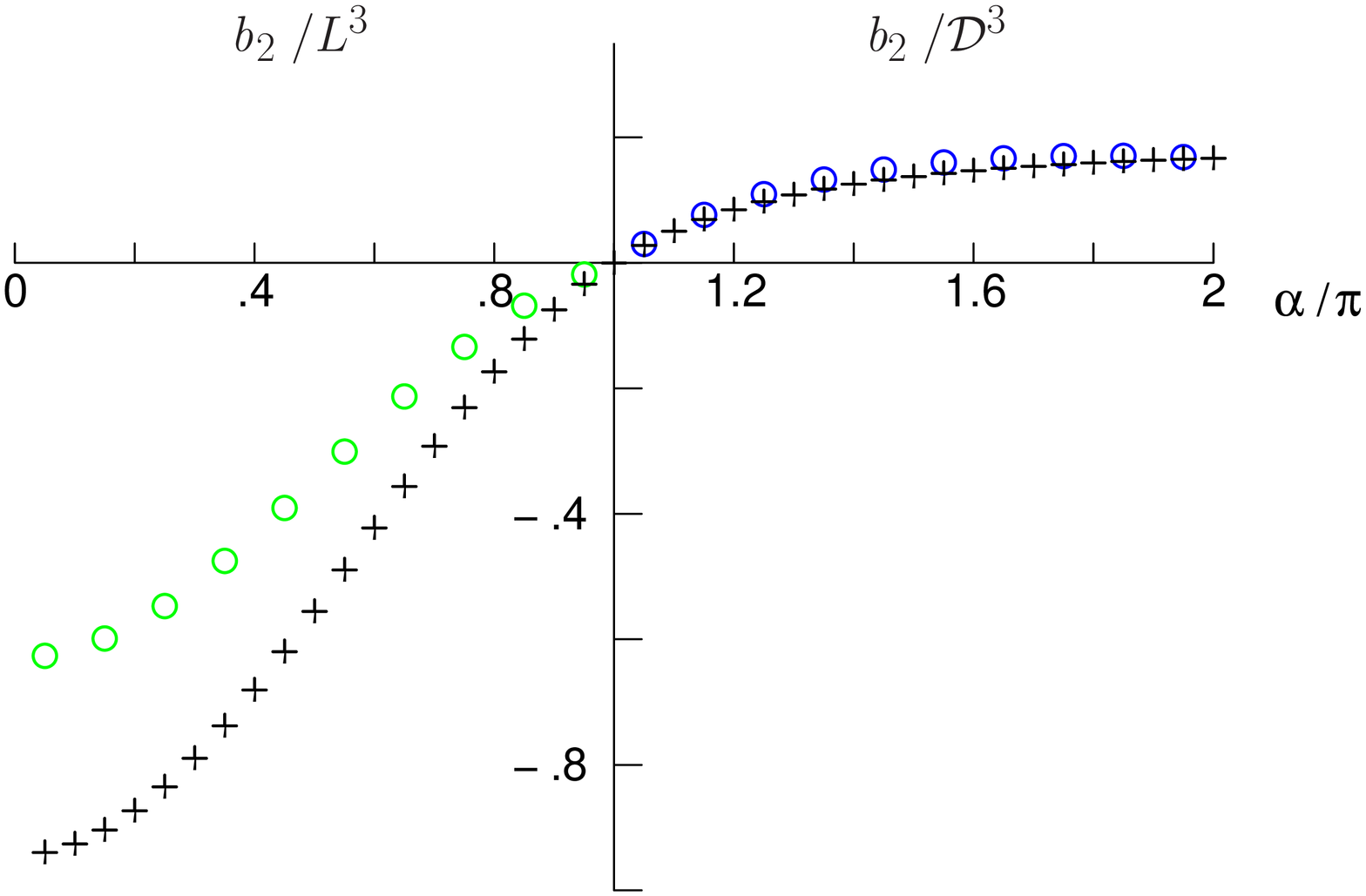}
%
%
\caption{\label{fig5} The anisotropy amplitude $b_2$ in
(\ref{small}).}
\end{figure}

\begin{thebibliography}{99}
%
\bibitem{rudh98}D. Rudhardt and C. Bechinger and P. Leiderer, Phys. Rev.
Lett. {\bf 81}, 1330 (1998).

\bibitem{manoha}A. van Blaaderen,
Science {\bf 301}, 470 (2003); MRS Bulletin, February, 85 (2004).

\bibitem{snoeblaa}E. Snoeks et al.
Adv. Mater. {\bf 12}, 1511 (2000).

\bibitem{ebm03}
E. Eisenriegler, A. Bringer, and R. Maa{\ss}en. J. Chem. Phys.
{\bf 118}, 8093 (2003).
%
\bibitem{PCI}E. Eisenriegler, in {\em Soft
Condensed Matter}, (G. Gompper and M. Schick eds.), Wiley-VCH
(2005).
%
\bibitem{gen79}P.G. de Gennes, {\em Scaling Concepts in Polymer Physics}
(Cornell University, Ithaca, 1979).
%
\bibitem{be95}T. Burkhardt and E. Eisenriegler, Phys. Rev. Lett. {\bf 74},
3189 (1995); {\bf 78}, 2867 (1997).
%
\bibitem{er95}E. Eisenriegler and U. Ritschel, Phys. Rev. B{\bf 51},
13717 (1995).
%
\bibitem{eise04}E. Eisenriegler, J. Chem. Phys. {\bf 121}, 3299 (2004).
%
\bibitem{contact}First, Eq. (\ref{small}) applies to correlation
functions with operators much further away from ${\bf r}_{\rm P}$
than the particle size. Apart from contact-terms \cite{ebm03}, Eq.
(\ref{small}) may also be used for calculating integrals such as
(\ref{inter}). Here we only consider the leading isotropic and
anisotropic contributions for small particle size, and the contact
terms do not contribute. The leading contact term has particle
size exponent $d$ and is isotropic \cite{ebm03}.
%
\bibitem{transl}Compare the discussion in Refs.
\onlinecite{ebm03,PCI} for the Gaussian model. In the notation of
Ref. \onlinecite{PCI} with amplitudes $\beta$ in Eqs. (162-165),
$a_1 \sqrt{B_\epsilon}=\beta_{\rm I}\tilde{S}_d/\sqrt{2} , \, b_1
\sqrt{B_\epsilon}= (\beta_{\rm VI}(d-2)+\beta_{\rm VII}d)
\tilde{S}_d/(4\sqrt{2}(d-1)), \, b_2=(-\beta_{\rm VI}+\beta_{\rm
VII})/2$ with $\tilde{S}_d$ given in (\ref{phi2wedge}).
%
\bibitem{cardy}J.L. Cardy in {\it Phase Transitions and Critical
Phenomena}, edited by C. Domb and J.L. Lebowitz (Academic Press,
London, 1986), Vol. {\bf 11}, p. 55.
%
\bibitem{be85}T. Burkhardt and E. Eisenriegler, J. Physics A {\bf
18}, L83 (1985).
%
\bibitem{hksd99}A. Hanke, M. Krech, F. Schlesener. and S.
Dietrich, Phys. Rev. E {\bf 60}, 5163 (1999).
%
\bibitem{edgenote}While we concentrate in Secs. IV-VI on
nonadsorbing ideal chains represented by the Gaussian model with
Dirichlet boundary conditions, the discussion in Sec. III does not
specify the bulk and surface universality classes \cite{cardy} and
allows also for other applications \cite{eise04,hksd99}.
%
\bibitem{cardyfilm}J.L. Cardy, Nucl. Phys. B{\bf 290}, 355 (1987).
%
\bibitem{mao95}D.M. Mc Avity and H. Osborn, Nucl. Phys. B{\bf
455}, 522 (1995).
%
\bibitem{buxu91}(a) T.W. Burkhardt and T. Xue, Phys. Rev. Lett.
{\bf 66}, 895 (1991); (b) Nucl. Phys. B{\bf 354}, 653 (1991); (c)
T.W. Burkhardt and I. Guim, Phys. Rev. B{\bf 36} 2080 (1987); (d)
Phys. Rev. B{\bf 47} 14 306 (1993).
%
\bibitem{bc80}(a) L.S. Brown, Ann. Phys. (N.Y.) {\bf 126}, 135 (1980); (b) L.S.
Brown and J.C. Collins, ibid. {\bf 130}, 215 (1980).
%
%
\bibitem{cardywedge}J.L. Cardy, J. Phys. A{\bf 16}, 3617 (1983).
%
\bibitem{examp}For $\alpha=\pi/g$, where $\delta \langle
\hat{\varphi}_{12} \rangle_{\rm wedge}$ is a superposition of
images, $\bar{\cal E}$ is an entire function of $d$. For example,
$\bar{\cal E}$ is given by $(2 {\rm cos} \Omega)^{2-d} /\sqrt{2}$
if $\alpha=\pi$, and by (\ref{engypi2}) if $\alpha=\pi/2$. For
$\alpha$ arbitrary, the analytic continuation of
$(\sqrt{2}/\alpha) \, {\rm sin} (\pi d/2) J^{(2)}$ with $J^{(2)}$
from (\ref{inte0''}), and thus of $e_0$, are analytic functions of
$d$ in the interval $2-(2\pi/\alpha)<d<4$. The continuation of
$e_0$, after taking the limit in (\ref{limdelprop}) below $d=2$,
coincides with the result from taking the limit above $d=2$
directly, since both certainly coincide for $\alpha=\pi/g$ and are
analytic in $\alpha$. Similar statements apply to $J_{\rm e}$ in
(\ref{Phiebeta3}). Here the continuation of $J_{\rm
e}/\tilde{S}_d$ is an analytic function of $d$ for
$-(2\pi/\alpha)<d<4$.
%
\bibitem{bey}The contributions $3 \beta_{\rm I}^2 \Phi^4 /4!$ and $-15\beta_{\rm I}^3 \Phi^6
/6!$ of the nonleading isotropic operators $\Phi^4$ and $\Phi^6$
to $\sigma_I$ on the right hand side of Eq. (\ref{small'}) follow
from the relations $\langle \Phi^4 \rangle_{\rm particle}=
3\langle \Phi^2 \rangle_{\rm particle}^2$ and $\langle \Phi^6
\rangle_{\rm particle}= 15\langle \Phi^2 \rangle_{\rm particle}^3$
between profiles due to Wick's theorem. Here $\beta_{\rm I}$
equals $a_1 \sqrt{B_\epsilon} \sqrt{2} / \tilde{S}_d$ and appears
in the leading isotropic contribution $a_1 \epsilon=-\beta_{\rm I}
\Phi^2 /2$ in the notation of Ref. \onlinecite{PCI}. For the
sphere with $\alpha=\pi$ and $\beta_{\rm I}=(L/2)^{d-2} /
\tilde{S}_d$ and the dumbbell of two touching spheres with
$\alpha=0$ and $\beta_{\rm I}=(L/2)^{d-2}
2(1-2^{3-d})\zeta(d-2)/\tilde{S}_d$ the role of these nonleading
operators has been discussed in Refs. \onlinecite{ebm03} and
\onlinecite{eise04}, respectively.
%
\bibitem{fxi}For the circumscribing ellipsoids in Eqs. (\ref{prolat})
and (\ref{oblat}), the parameters $[f, \, \xi_{\rm E}]$ of Ref.
\onlinecite{PCI} are given by $[ \, 2^{-1/2} L {\rm
cos}(\alpha/4)\sqrt{{\rm cos}(\alpha/2)} \, , \, 2^{1/2} /
\sqrt{1-{\rm tg}^2(\alpha/4)} \, ]$ and $[ \, ({\cal D} /2)
\sqrt{1-{\rm ctg}^2(\alpha/4)} \, , \, 1/ \sqrt{{\rm
tg}^2(\alpha/4) -1} \, ]$, respectively.
%
%
\bibitem{weak}For example from (\ref{weakprop}) one may calculate
$\langle \Phi^2({\bf r}) \rangle$ and $\langle T_{kl}({\bf r})
\rangle$, and from the behavior for $R \ll r$ obtain $a_1, \,
b_1$, and $b_2$.
%
\end{thebibliography}
\end{document}